\documentclass[a4,letters,usenatbib]{mnras}

\usepackage{amsmath}
\usepackage{amssymb} 
\usepackage{multirow}
\usepackage{textcomp}
\usepackage{microtype}
\usepackage{subcaption}
\usepackage[dvipsnames]{xcolor}
\usepackage{ulem}
\usepackage{graphicx}
\usepackage{subcaption}
\usepackage{float}
\usepackage{mathptmx}
\usepackage{tabularx}
\usepackage{caption}
\usepackage{acronym}
\usepackage{bm}
\usepackage[T1]{fontenc}

\newcommand{\skyvec}{\boldsymbol{\hat{\varOmega}}}
\newcommand{\hp}{h^{+}}
\newcommand{\hc}{h^{\times}}

\newcommand{\Ap}{A^{+}}
\newcommand{\Ac}{A^{\times}}

\newcommand{\Mmat}{\bm{\mathsf{M}}}

\newcommand{\loc}{\varOmega_{90}}
\newcommand{\nfifty}{N_{50}}
\newcommand{\nninety}{N_{90}}
\newcommand{\mmbulge}{M-M_\mathrm{b}}
\newcommand{\mbulge}{M_\mathrm{b}}
\newcommand{\mchirp}{\mathcal{M}}
\newcommand{\dlum}{D_\mathrm{l}}

\newcommand{\msun}{\textrm{M}_\odot}

\usepackage[T1]{fontenc}
\usepackage{ae,aecompl}

\title[PTA Host Galaxy Candidates]{Associating Host Galaxy Candidates to Massive Black Hole Binaries resolved by Pulsar Timing Arrays}

\author[Goldstein et al.]{Janna M. Goldstein$^{1}$\thanks{E-mail: \href{mailto:jgoldstein@star.sr.bham.ac.uk}{jgoldstein@star.sr.bham.ac.uk}},
  Alberto Sesana$^{1}$,
  A. Miguel Holgado$^{2}$ and
	John Veitch$^{1,3}$
\\
$^{1}$School of Physics and Astronomy and Institute for Gravitational Wave Astronomy,\\ University of Birmingham, Edgbaston, Birmingham, B15 2TT, UK \\
$^{2}$Department of Astronomy and National Center for Supercomputing Applications,\\ University of Illinois at Urbana-Champaign, Urbana, IL, 61801, USA \\
$^{3}$School of Physics and Astronomy, University of Glasgow, Glasgow, G12 8QQ, UK
}
\date{Last updated \today; in original form \today.}

\pubyear{2018}

\graphicspath{{./figures/}}

\begin{document}
\label{firstpage}
\pagerange{\pageref{firstpage}--\pageref{lastpage}}
\maketitle

\begin{abstract}
  We propose a novel methodology to select host galaxy candidates of future pulsar timing array (PTA) detections of resolved gravitational waves (GWs) from massive black hole binaries (MBHBs). The method exploits the physical dependence of the GW amplitude on the MBHB chirp mass and distance to the observer, together with empirical MBH mass--host galaxy correlations, to rank potential host galaxies in the mass--redshift plane. This is coupled to a null-stream based likelihood evaluation of the GW amplitude and sky position in a Bayesian framework that assigns to each galaxy a probability of hosting the MBHB generating the GW signal. We test our algorithm on a set of realistic simulations coupling the likely properties of the first PTA resolved GW signal to synthetic all-sky galaxy maps. For a foreseeable PTA sky-localization precision of 100\,deg$^2$, we find that the GW source is hosted with $50\%$($90\%$) probability within a restricted number of $\lesssim50$($\lesssim500$) potential hosts. These figures are orders of magnitude smaller than the total number of galaxies within the PTA sky error-box, enabling extensive electromagnetic follow-up campaigns on a limited number of targets.
\end{abstract}




\section{Introduction}
Multimessenger astronomy with gravitational waves (GWs) has long been anticipated as one of the `Holy Grails' for the understanding of the Universe. After a long wait, the first spectacular confirmation of its potential came with the detection of GWs from GW170817 \citep{2017PhRvL.119p1101A}, a binary neutron star coalescence (BNS) at about 40\,Mpc distance, accompanied by a bright electromagnetic (EM) signal observed at all wavelengths \citep{2017ApJ...848L..12A}. The wealth of fresh information brought by this event has been key to confirming several theoretical speculations, from the short gamma ray burst--BNS merger connection \citep{2017ApJ...848L..13A}, to the synthesis through r-processes of the heavy elements permeating the Universe \citep{2017ApJ...848L..19C}, and opened a new way to do cosmology with standard sirens \citep{schutz_determining_1986,2017Natur.551...85A,Fishbach:2018gjp}. All of this has been achieved thanks to the excellent sky-localization and distance information provided by LIGO-Virgo, an intense follow-up campaign, and the presence of a bright, distinct EM counterpart that could be easily singled out from other possible candidates. The small size of the sky-localization error-box was crucial, since it allowed systematic scanning of a relatively low number of possible galaxy hosts.

Realizing the full potential of multimessenger astronomy might prove more difficult in the low frequency band relevant to space based interferometers such as LISA \citep{2017arXiv170200786A} and pulsar timing arrays \citep[PTAs][]{2016MNRAS.458.1267V}, where the expected loudest sources involve inspiral and merger of massive black hole binaries (MBHBs) at cosmological distances \citep{2008MNRAS.390..192S,2016PhRvD..93b4003K}. Merging MBHBs are not {\it per se} expected to produce EM signals, so multimessenger efforts need to rely on some distinctive signature in the emission of the gas that might be accreted by the system during the inspiral and final coalescence \citep{doi:10.1093/mnras/stx1130}. Even so, it is not clear what that signature would be, and a range of possibilities have been proposed, from periodicity \citep[e.g.][]{2012MNRAS.420..860S} to peculiar spectral features \citep[e.g.][]{tanaka_electromagnetic_2012} and electromagnetic chirps \citep[e.g.][]{haiman_electromagnetic_2017}.

The situation is particularly challenging for PTAs. Besides detecting a stochastic gravitational wave background (GWB) produced by the superposition of many MBHB systems \citep[e.g.][]{Phinney2001, 2008MNRAS.390..192S,2012ApJ...761...84R}, PTAs also have the capability to detect and localize in the sky particularly loud MBHBs \citep{2009MNRAS.394.2255S, 2015MNRAS.447.2772R, 2015MNRAS.451.2417R, 2018MNRAS.477..964K}.  \citet{2017NatAs...1..886M} predict that in 10 years, the first resolved binary could be detected. Strategies for optimizing PTA for single source detection (by allocating observing time and targeted searched for new pulsars) have been proposed by e.g. \citet{2011ApJ...730...17B}, \citet[by identifying "hot spots" from nearby galaxy clusters]{2014ApJ...784...60S} and \citet{2018ApJ...868...33L}.
However both the prediction of \cite{2018MNRAS.477..964K} and the optimization for a detection of \cite{2018ApJ...868...33L} are complicated by the difficult to model red noise of the pulsars.

When looking for an EM counterpart to the first PTA resolved binary detections,
one faces three main problems. First, the sky-localization is expected to be relatively poor \citep[of the order of hundreds of deg$^2$][]{2010PhRvD..81j4008S,Goldstein:2017qub}. Second, the detected GW signal is likely to be monochromatic. The absence of observable frequency evolution (chirp) of the waveform prevents one from separating the source mass from the distance, since only the overall amplitude $A$ and frequency $f$ are measured.
Last, the signal evolves slowly in time, with a periodicity of the order of years. Associated counterparts might be identified through peculiar features in the source luminosity or through potential peculiarities of the galaxy host \citep{2012MNRAS.420..860S,tanaka_electromagnetic_2012,2013CQGra..30v4013B} In any case there is no clear smoking-gun event such as a transient counterpart, as is the case for a BNS merger.

It is therefore crucial to find a way to identify the most promising host galaxy candidates among the millions of objects falling within the source sky location error-box. In this paper, we develop a Bayesian framework to identify the most likely hosts by matching the information contained in a hypothetical PTA detection to candidate galaxy properties. The key point around which our analysis is built, is that individually resolvable sources in the PTA band necessarily have a large strain amplitude $A$ \citep{2015MNRAS.451.2417R, 2018MNRAS.477..964K}, which can result only from particularly massive and/or nearby MBHBs. We show that this allows one to exclude at high confidence the vast majority of the galaxies in the error-box, significantly reducing the number of candidates.

To demonstrate this, we consider a synthetic PTA and inject GW signals with properties compatible to the first single sources to be detected by future PTAs, drawn by following the procedure described in \citet{2015MNRAS.451.2417R}.
We then use the null-stream analysis developed in \citet{Goldstein:2017qub} to construct the 3-D likelihood function of the signal amplitude $A$ and sky-localization $\theta, \phi$. We extract a mock catalog of galaxies from the synthetic all sky maps obtained by \cite{2012MNRAS.421.2904H} from the Millennium Simulation \citep{2005Natur.435..629S} and we use Bayesian inference to rank host candidates.

The paper is organized as follows. In Section \ref{sec:meth} we lay out the mathematical basis of our experiment, including the construction of a likelihood from null-streams and the Bayesian framework for the computation of a host galaxy probability. This framework is then applied in Section \ref{sec:setup} to a number of representative simulations  with results laid out in Section \ref{sec:res} and the main conclusions and outlook presented in Section \ref{sec:conc}.

\section{Mathematical framework} \label{sec:meth}
\subsection{Signal model and null-stream sky-localization}
PTAs are capable of reconstructing the incoming direction of a deterministic GW source via triangulation \citep{2012PhRvD..86l4028B,2012PhRvD..85d4034B}, providing that three or more millisecond pulsars (MSPs) contribute to the detection. We consider, for simplicity, a circular, monochromatic MBHB. The emitted GW can be written in the form \citep{1998PhRvD..58f3001J}
\begin{align}
	\hp(t) &= \Ap \cos{(2\psi)} - \Ac \sin{(2\psi)}
	\label{eq:general_hp} \\
	\hc(t) &= \Ap \sin{(2\psi)} + \Ac \cos{(2\psi)},
	\label{eq:general_hc}
\end{align}
where $\psi$ is the GW polarization angle and
\begin{align}
\Ap &= A \frac{1}{2} (1 + \cos{\iota}^2) \cos{(2\pi f t + \phi_0)} \label{eq:polarisation_Ap} \\
\Ac &= A (\cos{\iota}) \sin{(2\pi f t + \phi_0)}.
\label{eq:polarisation_Ac}
\end{align}
The two polarization amplitudes $\Ap,\Ac$ are modulated with the {\it observed} GW frequency $f$ and are related to the intrinsic amplitude \footnote{This definition of $A$ is equivalent to the definition with a prefactor of 2 instead of 4 - which is also seen in the literature e.g. in \citet{doi:10.1093/mnras/stv2092} - as that definition is accompanied by an additional factor of 2 in Equations \ref{eq:polarisation_Ap}, \ref{eq:polarisation_Ac}.}
\begin{equation}
  A=4\frac{(\mathrm{G}{\cal M}_z)^{5/3}(\pi f)^{2/3}}{\dlum}
  \label{eq:A}
\end{equation}
via the inclination angle to the line of sight $\iota$. The amplitude $A$ is a function of the source redshifted chirp mass
\begin{equation}
  {\cal M}_z=(1+z){\cal M}=(1+z)\frac{(M_1M_2)^{3/5}}{(M_1+M_2)^{1/5}},
  \label{eq:Mz}
\end{equation}
and of its luminosity distance 
\begin{equation}
  \dlum=(1+z)D_\mathrm{H}\int_0^z \frac{dz'}{E(z')}.
  \label{eq:dl}
\end{equation}
In the above equations, $M_1$ and $M_2$ are the masses of the two black holes forming the binary, $z$ is the source redshift, $D_\mathrm{H}=c/H_0$ and $E(z)=\sqrt{\Omega_\mathrm{M}(1+z)^3+\Omega_\mathrm{\Lambda}}$, with $\Omega_\mathrm{M}$ and $\Omega_\mathrm{\Lambda}$ being the fractional mass and cosmological constant energy content, $H_0$ the Hubble constant and assuming a standard flat $\Lambda$CDM Universe \citep{2016A&A...594A..13P}.

The GW induces into the pulse time of arrival a redshift of the form
\begin{equation}
  z(t, \skyvec) =  F^+ (\skyvec) \hp(t) + F^{\times} (\skyvec) \hc(t)
  \label{eq:response_functions}
\end{equation}
where the `antenna beam patterns' $F^+$ and $F^{\times}$ depend on the angle between the incoming GW directrion $\skyvec$ and the known position of the MSP \citep[see e.g.][]{anholm09}. In practice, PTAs are sensitive to the two wave polarizations $\hp,\, \hc$ that depend on the vector of parameters $(A,\iota,f,\psi,\phi_0,\theta,\phi)$, where we decomposed the incoming wave direction $\skyvec$ onto its ($\theta,\phi$) coordinates in the sky.\\

In \citet{Goldstein:2017qub} we developed a null-stream based analysis (see also  \citet{2015MNRAS.449.1650Z} and \citet{2016arXiv160703459H}) that, among other things, can be used to infer the amplitude and incoming direction of the GW source. Since for an individual GW source there are only two polarizations, but an array of $N$ allows measurement of $N$ independent time (or frequency) series, it is possible to apply a matrix transformation that `collapses' the signal into two of these time series. This nulls the signal contribution in all the others, hence constructing $N-2$ null-streams. Formally, the transformation takes the form \citep[see][ for details]{Goldstein:2017qub}:
\begin{equation}
\Mmat\, \bm{d} = \begin{pmatrix}
	\hp \\
	\hc \\
	\eta_1 \\
	\vdots \\
	\eta_{N-2}
\end{pmatrix} + \Mmat\, \bm{n} 
\equiv \bm{h} + \Mmat\, \bm{n},
\label{eq:strainspace_data}
\end{equation}
where $\bm{d}$ represents the original $N$ time series of the $N$ pulsars (including signal and noise $\bm{n}$), $ \Mmat$ is the matrix transformation, $\eta_i = 0$ are the null streams, and $\bm{h}$ is the combined vector of GW polarisations and null streams. In practice this amounts to the construction of $N$ linear combinations of the timing residuals so that the GW signal is present only in two of them and null in all the others.

\begin{figure}
	\begin{subfigure}{\columnwidth}
		\includegraphics[width=\columnwidth]{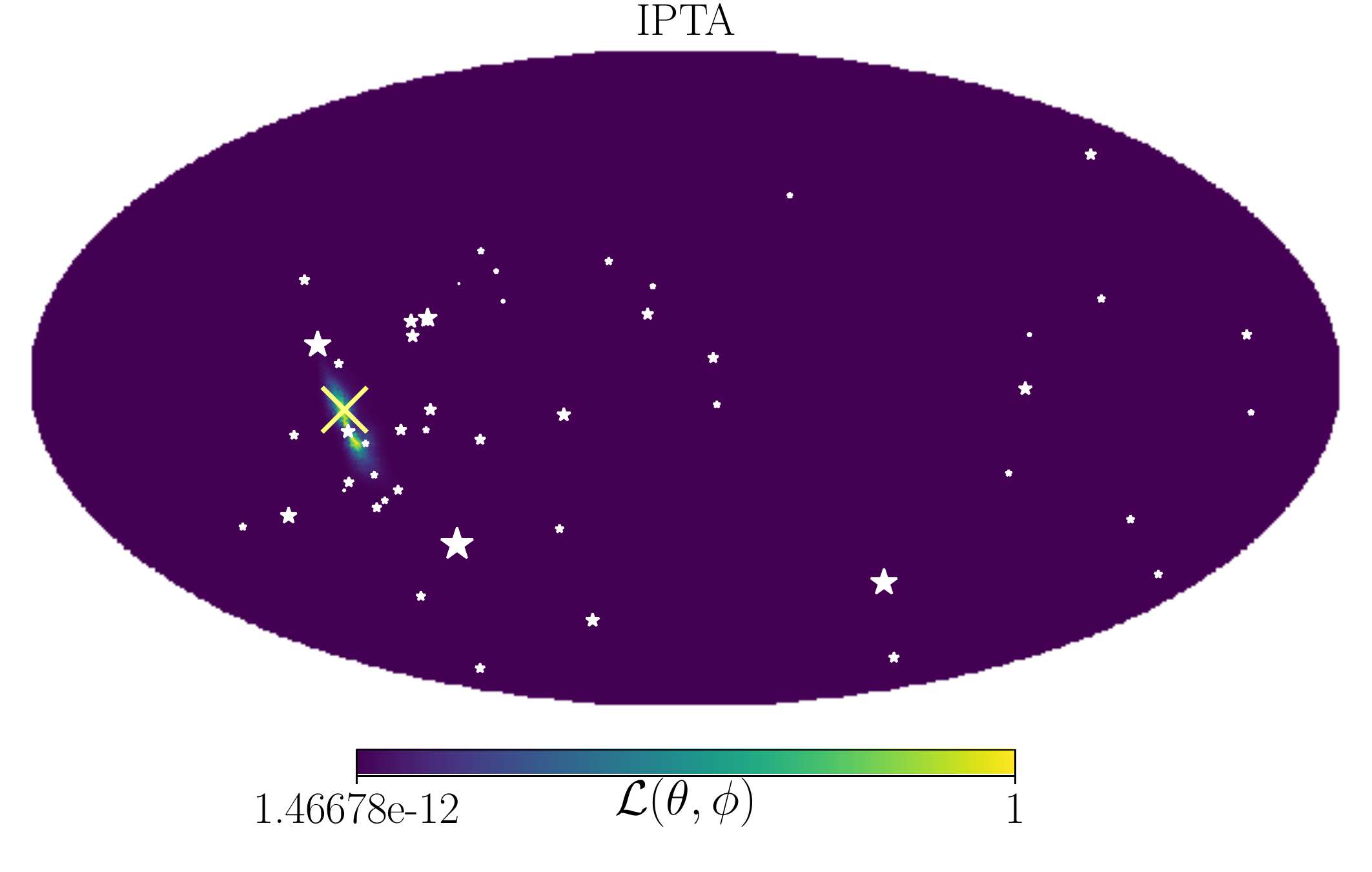}
		\label{fig:L_skymap}
	\end{subfigure} 
	\newline
	\begin{subfigure}{\columnwidth}
		\includegraphics[width=\columnwidth]{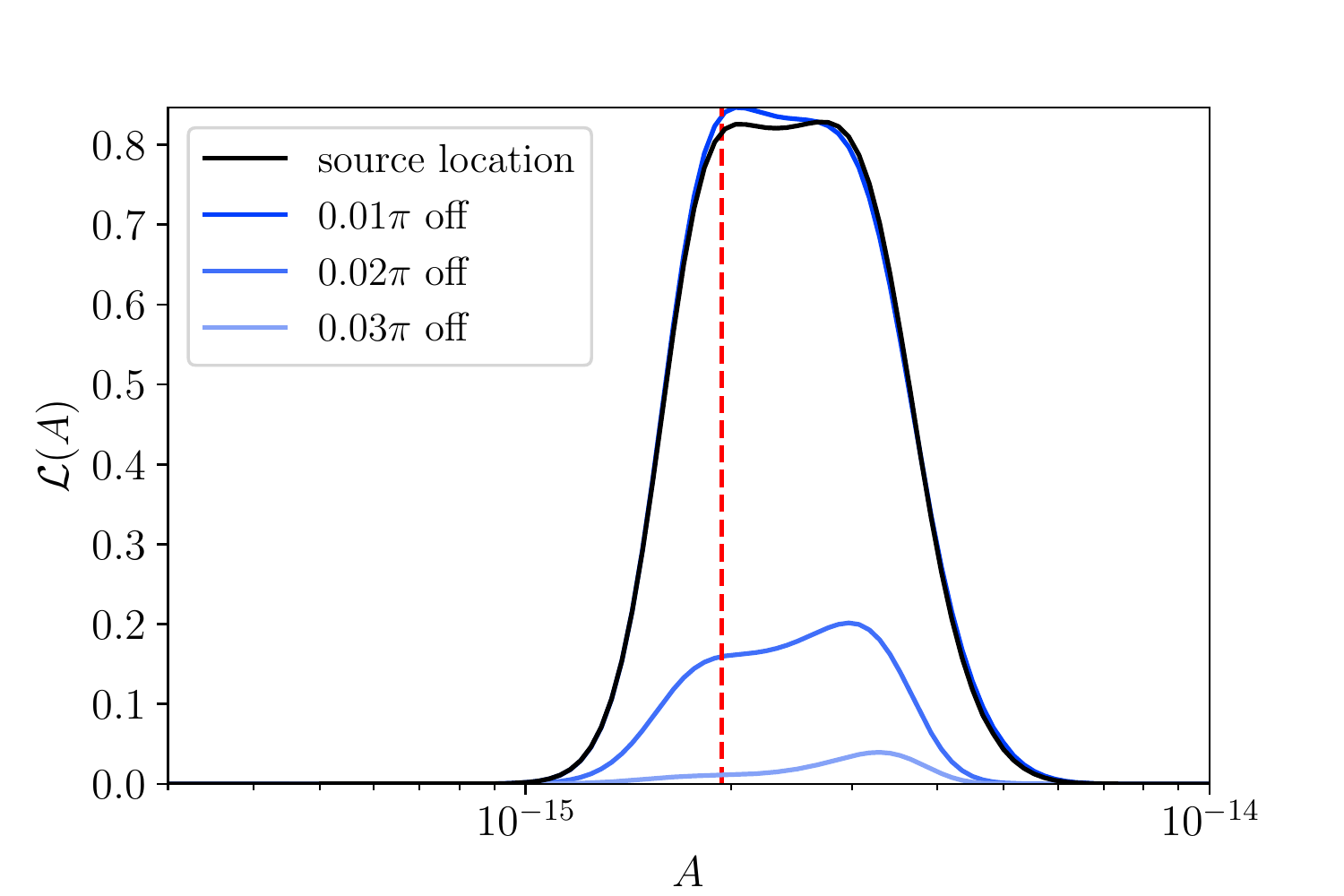}
		\label{fig:L_amps}
	\end{subfigure}
	\caption{Example of ${\cal L}(A,\theta,\phi)$ as output by the null-stream pipeline. The injected signal is for source A, with $S/N = 12$ (see Section \ref{sec:setup}). Top: Likelihood marginalized over $A$ (i.e. ${\cal L}(\theta,\phi)$ with an arbitrary normalization). The IPTA pulsars are marked with stars, where the size of the star corresponds to the noise level of the pulsar (with bigger stars for lower noise). The yellow cross indicates the position of the injected source.
	Bottom: ${\cal L}(A|\theta_s, \phi_s)$ at the source position $(\theta_s, \phi_s)$ (in black) and at some offset positions $(\theta_s, \phi_s + \Delta)$ (in blue). The likelihoods are normalized only with respect to each other. The red dashed line is placed at the injected amplitude value.}
	\label{fig:L_example}
\end{figure}

The null streams can then be used to construct the likelihood function
\begin{equation}
l = -\frac{1}{2} \bigg( (\Mmat\bm{d} - \bm{h})^{\top} ((\Mmat^{-1})^{\top} 
\Gamma \Mmat^{-1}) (\Mmat\bm{d} - \bm{h}) \bigg) + norm.
\label{eq:likelihood}
\end{equation}
where $\Gamma$ is the inverse of the covariance matrix appropriate for the expected noise of the detector. For a signal detected at frequency $f$, marginalization of the likelihood over the parameters $\iota, \psi, \phi$, yields the 3-D likelihood function ${\cal L}(A,\theta,\phi)$. For an example ${\cal L}(A,\theta,\phi)$, see Figure \ref{fig:L_example}.

In this work, we use the \citet{Goldstein:2017qub} null-stream pipeline to obtain ${\cal L}(A,\theta,\phi)$. However in principle any method could be used to localize the source, as long as it can provide a joint likelihood on the sky location and amplitude of the signal. The framework for candidate host galaxy selection which is introduced in the following section, is written in term of a generic input ${\cal L}(A,\theta,\phi)$.

\subsection{Bayesian inference for galaxy host} \label{sec:Bayes}
Our goal is to combine the likelihood information ${\cal L}(A,\theta,\phi)$ with individual galaxy properties to assess the probability of each given galaxy to be the host of the detected GW source. The question we want to answer in practice is: given the detection of a signal with 3-D likelihood described by ${\cal L}(A,\theta,\phi)$, what is the probability that a galaxy $G_i$ described by a set of observed parameters $\bm{\lambda}$ -- known with prior probability $p(\bm{\lambda}|G_i)$ -- is the host of the signal source? To answer this question we need a theoretical model that connects the strength and location of a putative GW signal to observable galaxy parameters.

Since MBHBs reside in the center of galaxies, the sky coordinates of each specific galaxy $(\theta_G,\phi_G)$ coincide with the sky coordinates of the putative GW source. We therefore have $\theta_G=\theta$ and $\phi_G=\phi$. Furthermore, we see from equations (\ref{eq:A}) and (\ref{eq:Mz}) that the GW amplitude $A$ depends on the source chirp mass $\mchirp$, and luminosity distance $\dlum$. This latter can be easily measured from the galaxy spectroscopic redshift by assuming a fiducial cosmology. Whereas $\mchirp$ can be written in terms of the total binary mass $M$, and mass ratio $q=M_2/M_1$ (with $M_2 \leq M_1$) as: $\mchirp=Mq^{3/5}/(1+q)^{6/5}$. We can assume the total mass to be related to the bulge mass via an $\mmbulge$-relation of the form 
\begin{equation}
  {\rm log}_{10}\left(\frac{M}{\msun}\right) = \alpha+\beta{\rm log}_{10}\left(\frac{\mbulge}{10^{11}\msun} \right) \ ,
  \label{eq:MMb}
\end{equation}
which connects the total binary mass to the observable galaxy bulge stellar mass $\mbulge$.
If we group the $\mmbulge$ constants $\alpha$ and $\beta$ with the galaxy parameters, the vector of seven parameters
\begin{equation}
\bm{\lambda}=(\mbulge, \dlum,\theta, \phi, q, \alpha, \beta)
\label{eq:parameter_vector},
\end{equation}
is sufficient to connect a specific galaxy to the GW strain. All of them but $q$, $\alpha$ and $\beta$ can be directly extracted from observations. \\

Formally the full calculation can be cast in term of Bayes' theorem. Let $P(G_i|d)$ be the probability of galaxy $G_i$ being the host galaxy, given some data $d$, then:
\begin{equation}
P(G_i |d) = \frac{P(G_i)}{P(d)} P(d|G_i) = \frac{P(G_i)}{P(d)} \int p(d| \bm{\lambda}) p(\bm{\lambda}|G_i) \ {\rm d}\bm{\lambda},
\label{eq:prob}
\end{equation}
where $P(d) =\sum_i P(d|G_i)$ is the likelihood of the data marginalized over all galaxies (or evidence). $P(G_i)$ is the prior probability of $G_i$ being the host, which we take to be a constant, having no reason \textit{a priori} to prefer any particular galaxy. Of interest is the shape of the distribution of $P(G_i|d)$, so disregarding the constant prefactor ${P(G_i)}/{P(d)}$, we are left with the likelihoods $P(d|G_i)$.

The likelihood of a specific galaxy $G_i$ to be the host of the GW source is given by the integral in equation (\ref{eq:prob}) and is composed of the probability of the data given the source parameters $p(d| \bm{\lambda})$, times the prior distribution on these parameters $p(\bm{\lambda}|G_i)$, integrated over all the relevant variables given in Equation \ref{eq:parameter_vector}.

What is needed is an operational form for $p(d| \bm{\lambda})$. First, the amplitude $A$ is independent on $\theta,\phi$ and so:
\begin{equation}
p(d|\bm{\lambda}) = 
p(d|\mbulge, \dlum, \theta, \phi, q, \alpha, \beta) = p(d|A, \theta, \phi)\, p(A|\mbulge, \dlum, q, \alpha, \beta).
\end{equation}
Second, $A$ is a direct function of the chirp mass ${\cal M}$ and distance only, we can therefore write
\begin{equation}
p(A|\mbulge, \dlum, q, \alpha, \beta) = p(A|{\cal M}, \dlum)\, p({\cal M}|\mbulge, q, \alpha, \beta). 
\end{equation}
Last, ${\cal M}$ is a function of $q$ and $M$, and the latter is related to $\mbulge$ by the $\mmbulge$-relation. We therefore have
\begin{equation}
p({\cal M}|\mbulge, q, \alpha, \beta) = p({\cal M}|M, q)\, p(M|\mbulge, \alpha, \beta).
\end{equation}
Putting the chain together we get:
\begin{multline}
  p(d|G_i) =\int p(d|A, \theta, \phi)\, p(A|{\cal M}, \dlum)\, p({\cal M}|M, q)\, p(M|\mbulge,  \alpha, \beta) \\ p(\mbulge, \dlum, \theta, \phi, q, \alpha, \beta|G_i) \ {\rm d}\mbulge \, {\rm d} \dlum \, {\rm d}\theta \, {\rm d}\phi \, {\rm d}q \, {\rm d} \alpha \, {\rm d}\beta \, {\rm d}M .
  \label{eq:pgal}
\end{multline}

We can now specify the individual elements of equation (\ref{eq:pgal}) for practical computational purposes. 

\begin{itemize}
\item $p(\bm{\lambda}|G_i) = p(M_b, \dlum, \theta, \phi, q, \alpha, \beta|G_i)$ describes the prior knowledge of each galaxy property and the underlying $\mmbulge$ constants. We assume that all five galaxy parameters -- so excluding $\alpha$ and $\beta$ -- are independent so that the prior can be factorized as $p(\bm{\lambda}|G_i)=\prod_{j=1}^5p(\lambda_j|G_i)$. In particular:
\begin{itemize}
\item  $\mbulge$ in real surveys is generally obtained from the galaxy luminosity via bulge-disk decomposition. $\mbulge$ is then computed from the bulge luminosity by assuming a stellar mass function. Typical uncertainties in this procedure can be up to a factor of two \citep{2009MNRAS.394..774L}. Nonetheless, as a first approximation, we take $\mbulge$ to be known exactly, reducing the prior $p(\mbulge)$ to a delta function (so the integral over $\mbulge$ drops out).
\item  $\dlum$ is computed from the spectroscopic redshift of the galaxy $z$ via equation (\ref{eq:dl}). Uncertainties on the cosmological parameters $H_0,\Omega_\mathrm{M},\Omega_\Lambda$ are of the order of a few percent \citep{2016A&A...594A..13P} and weak lensing is subdominant for the $z<1$ galaxies relevant here \citep{2010MNRAS.404..858S}. We therefore also assume $\dlum$ to be known exactly, reducing the prior $p(\dlum)$ to a delta function, dropping the integration over $\dlum$ from the likelihood marginalization.
\item $\theta,\phi$ are generally determined with arcsecond precision, which for any practical purposes can be treated as delta functions as well.
\item $q$, the binary mass ratio, is essentially undetermined. We therefore use a broad log flat prior between $-2 \leq \log_{10}(q) \leq 0$ (i.e. $0.01 \leq q \leq 1$).
\end{itemize}  
The impact of changing the adopted priors in the calculation are discussed in Section \ref{sec:disc}.

\item $p(d|A,\theta,\phi)$ is directly proportional to the likelihood in the 3-D amplitude-sky location space ${\cal L}(A,\theta,\phi)$ returned as a numerical function with finite resolution by our null stream based parameter estimation pipeline. Given the values of $A$, $\theta$ and $\phi$ from the priors, we select the numerical value from the sky pixel at $(\theta, \phi)$ and the closest sampled amplitude to $A$. The sampling range ($10^{-17}$--$10^{-14}$) is big enough to cover the area of interest, so for values of $A$ outside this range, the likelihood is set to zero.

\item $p(A|{\cal M}, \dlum)$ is determined by the GW quadrupole formula. Given the system chirp mass and distance, the amplitude is univocally determined by equation (\ref{eq:A}). We can thus write 
\begin{equation}
p(A|{\cal M}, \dlum)=\delta\left(A-4\frac{(\mathrm{G}{\cal M}_z)^{5/3}(\pi f)^{2/3}}{\dlum}\right).
\end{equation}

\item $p({\cal M}|M, q)$ is similarly computed from the mathematical definition of ${\cal M}$ in terms of $M,q$ as
\begin{equation}
p({\cal M}|M, q)=\delta\left({\cal M}-\frac{Mq^{3/5}}{(1+q)^{6/5}}\right).
\end{equation}

\item $p(M|\mbulge, \alpha, \beta)$ is a core ingredient of the calculation. The possibility of ranking galaxy hosts stems from the simple fact that extremely massive black holes are hosted in extremely massive galaxies, a relation that has to be handled with care. Once a specific $\mmbulge$ relation of the form given by equation (\ref{eq:MMb}) with intrinsic dispersion $\epsilon$ is given, the MBH total mass probability is described by a log-normal prior
\begin{equation}
  p(M|\mbulge)=\frac{1}{\sqrt{2\pi\epsilon^2}}\exp\left\{-\frac{\left[{\rm log}\frac{M}{\msun}-\left(\alpha+\beta{\rm log}\frac{\mbulge}{10^{11}\msun}\right)\right]^2}{2\epsilon^2}\right\},
\label{eq:pMMb}  
\end{equation}
that we integrate from $-3\epsilon$ to $+3\epsilon$ around the minimum and maximum  expectation values of $M$ (the range of $M$ values being due to the spread in $(\alpha, \beta)$).

The $\mmbulge$ relation is quite uncertain, as demonstrated by its many different flavors found in the literature. Using the compilation of $\mmbulge$ relations of \cite{2018NatCo...9..573M} we construct an observationally motivated prior distribution in $(\alpha,\beta)$ by the following procedure. We make many random draws of the pair $(\alpha,\beta)$ uniformly from the ranges $\alpha \in [7.63, 8.63]$ and $\beta \in [0.79, 2.14]$, and consider the pair valid if the resulting $\mmbulge$ line falls within the region enclosed by the compiled sample of relations in the range $10^6\msun<M<10^{10}\msun$. The resulting probability distribution $p(\alpha,\beta)$ is shown in figure \ref{fig:ABprior}. We then marginalize over the parameters $(\alpha, \beta)$ in the computation of $p(d|G_i)$ in Equation \ref{eq:pgal}. 
We assume $\epsilon=0.3$ throughout, which is the typical relation dispersion value reported in the literature. 

\begin{figure}
	\includegraphics[width=\columnwidth]{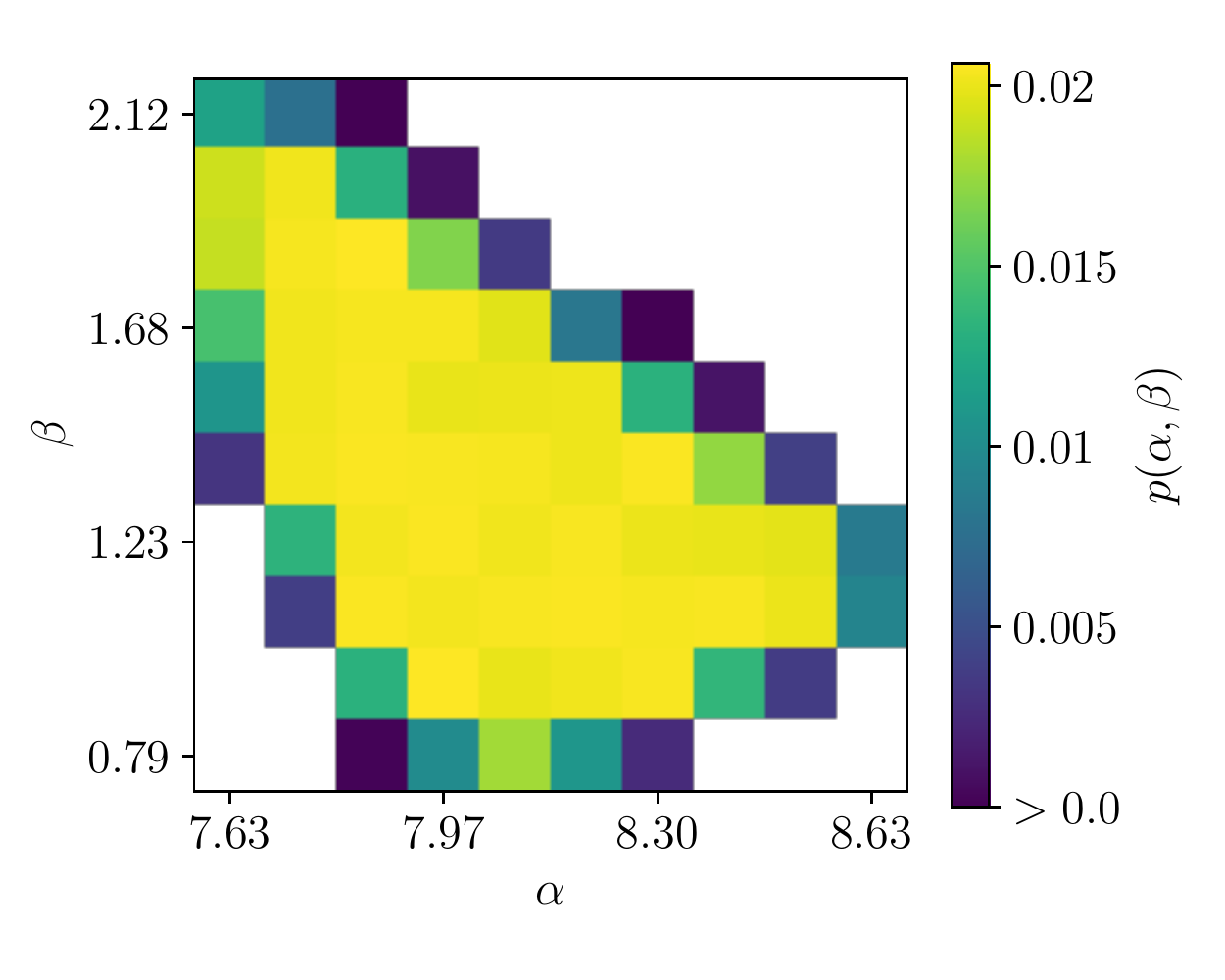}
	\caption{Prior on the $\mmbulge$ constants $(\alpha, \beta)$ constructed from the compilation of $\mmbulge$-relations in \citet{2018NatCo...9..573M}, see text in Section \ref{sec:Bayes}. The prior is binned in a $10\times10$ regular grid with $\alpha \in [7.63, 8.63]$ and $\beta \in [0.79, 2.14]$. The pixels are normalized such that their sum is one. Some combinations $(\alpha, \beta)$ have zero prior weight and are masked in white.}
	\label{fig:ABprior}
\end{figure}
\end{itemize}

Using these assumption, equation \ref{eq:pgal} reduces to a four dimensional integral over $M$, $q$, $\alpha$ and $\beta$. In the following, we show results where $\mmbulge$ is always marginalized over $(\alpha, \beta)$ and we discuss the impact of assuming a specific scaling relation in Section \ref{sec:disc}. In practice, we transform variables to $^{10}\log(M)$ and $^{10}\log(q)$ and perform the numerical integration in log space for these parameters.

\section{Practical implementation}
\label{sec:setup}

\subsection{Source selection}
To test our method, we simulate plausible future detections of single sources in PTA. We turn to work by \citet{2015MNRAS.451.2417R}, who have studied large scale simulations of MBHB populations and the resulting GW signals that could be detected by PTAs. They construct 20000 models (with different observed MBH mass functions, pair fractions and MBH-galaxy relations) and drew several Monte Carlo realizations of each model, to build realistic MBHB populations. They then considered the sensitivity of several PTAs as a function of time and used simple detection statistics to declare detection of either individual MBHBs or the overall stochastic background.  Although they find that it's more likely that the background is detected first, eventually, individual sources can also be confidently identified. For each of the simulations they record the properties of the first MBHB to be individually resolved by the PTA under consideration. Therefore, their procedure informs the likely parameters of the first resolvable MBHBs. We use it here to get the parameters for our test injections, as follows.

The signal-to-noise ratio (S/N) of a circular MBHB in an array of $M$ pulsars can be written as
\begin{equation}
\label{eq:snrs}
\text{S/N}=\left[ \sum_{i=1}^{M} (\text{S/N}_i)^2 \right]^{1/2},
\end{equation}
where the S/N in the $i$-th pulsar is
\begin{align}
\label{eq:snralpha}
(\text{S/N}_i)^2=\frac{A^2}{4\pi^2f^2S_i}{\cal R}(\vec\delta).
\end{align}
Here, $A$ is the GW amplitude given by equation \eqref{eq:A} and $f$ is the observed GW frequency. ${\cal R}(\vec\delta)$ is a factor of order unity that depends of the geometry of the system -- including source sky location and inclination, wave polarization angle and pulsar sky location -- and on the duration of the PTA observation $T$; see \cite{2015MNRAS.451.2417R} for the full expression. $S_i$ is the noise in the $i$-th pulsar which we consider to be of the form
\begin{equation}
\label{eq:nsd}
S_i=2\Delta t \sigma_i^2+S_{h,\text{rest}},
\end{equation}
where the first term on the rhs is the rms noise level of the timing residuals and the second term is the level of confusion noise given by all other sources contributing to the overall GW signal.

\begin{table}
	\centering
	\begin{tabular}{ccccc}
          \hline
          & & & & \\[-7 pt]
		source & ${\mathcal M}$\,[M$_\odot$] & z & $f$\,[nHz] & $ A $\\ 
          \hline
		A & $3.18\times10^9$ & 0.62 & 7.44 & $0.96\times10^{-15}$\\
		B & $5.36\times10^9$ & 0.57 & 5.94 & $2.05\times10^{-15}$\\
		C & $3.69\times10^9$ & 0.18 & 5.18 & $2.40\times10^{-15}$\\
          \hline
	\end{tabular}
	\caption{Properties of the three test sources selected for this study.}
	\label{tab:source_params}
\end{table}

To select suitable individual sources, we construct a mock version of the IPTA using the 49 pulsars of IPTA DR1 \citep{2016MNRAS.458.1267V}. We consider the actual sky location and rms noise $\sigma_i$ of each pulsar, and assume bi-weekly observations ($\Delta t=2$ weeks) for a timespan of $T=10$ years. Next, we generate 50 realizations of a realistic population of circular, GW driven MBHBs, based on one of the models presented in \cite{2013MNRAS.433L...1S}. The number of realizations is chosen to produce a sample of individually resolvable sources that is large enough to give us freedom to pick sources in desired regions in the sky (see below). In particular we use a fairly optimistic model resulting in a characteristic GW strain $h_{\rm c} = A(f/{\rm yr}^{-1})$ with $A\approx 1.3\times10^{15}$, which is just at the edge of the most recent PTA limits \citep{2015Sci...349.1522S,2015MNRAS.453.2576L,2016MNRAS.458.1267V,2018ApJ...859...47A}. 

In each model realization, we select the loudest GW sources one-by-one and use all remaining MBHBs to consistently compute $S_{h,\text{rest}}$. 
All potentially resolvable GW sources had S/N < 2 in the adopted setup.
This is a good sanity check for our simulation; in fact it is expected
that no observable sources result from this procedure, given that 
no single MBHB has been detected to date either.
To increase the $S/N$, we suppress the noise by multiplying each rms residual $\sigma_i$ by a fudge factor $\eta<1$. After decreasing $\eta$ to 0.2, 
we observe $\approx 30$ sources (in 50 GW signal realizations) at $S/N\gtrsim 5$. We select three of those sources, which we name A, B and C.

Relevant parameters of the selected sources are listed in Table \ref{tab:source_params} and their location in the sky, relative to the IPTA pulsars, can be seen in figure \ref{fig:localization_map}. We have intentionally picked three sources in areas of different IPTA pulsar density.
Because the response functions depend on the angular distance between the pulsar and the GW propagation direction (Equation \ref{eq:response_functions}), the localization behaviour is different for sources that are close to (good) pulsars than for those in relatively empty regions of the sky (see also Section \ref{sec:loc_results}).
Parameters listed in Table \ref{tab:source_params} are consistent with distributions shown in Figure 6 of \cite{2015MNRAS.451.2417R}. The first resolvable sources are likely to be at relatively low frequencies (few nHz) and can come from MBHBs at moderate redshifts (up to $z\approx 1$).

\subsection{Source injection and likelihood evaluation}

Each source is injected into a synthetic PTA, based on IPTA data release 1 \citep{2016MNRAS.458.1267V}. The sky location and relative white noise level for each pulsar are kept the same as in IPTA DR1 (see their Table 4 under Residual rms). Practical limitations on the method of \citet{Goldstein:2017qub} mean the cadence and observation time of each pulsar has to be the same, so these are averaged over. We adjust the total observation time and/or reduce the noise in each pulsar by a constant factor to set the $S/N$ of an injected source at the values 7, 10, 12, and 15 (see table \ref{tab:SNRadjustments}). We choose 7 as the smallest $S/N$ value because it ensures a confident detection according to the ${\cal F}$ statistic adopted by \cite{2015MNRAS.451.2417R} (and used in this work). Assuming a typical PTA and a false alarm probability of 0.001, a source with $S/N=7$ has a detection probability of $\approx 0.9$. For each setup, a likelihood ${\cal L}(A,\theta,\phi)$ is obtained using three different realizations of random white noise in the null stream pipeline. Summarizing, we run a total of 36 simulations featuring:
\begin{itemize}
\item {\it three} different sources: A, B, C;
\item {\it four} values of detection $S/N=$ 7, 10, 12, and 15;
\item {\it three} independent white noise realizations.\\
\end{itemize}
\begin{table}
	\centering
	\begin{tabular}{c|c||c|c||c|c||c|c}
		\hline
		\multicolumn{2}{r||}{}& \multicolumn{2}{c||}{} & \multicolumn{2}{c||}{} & \multicolumn{2}{c}{} \\[-7pt]
		\multicolumn{2}{r||}{source}& \multicolumn{2}{c||}{A} & \multicolumn{2}{c||}{B} & \multicolumn{2}{c}{C} \\
		\hline
		& & & & & & & \\[-7pt]
	
		\multirow{2}{*}{$S/N$}  & \multirow{2}{*}{\shortstack[l]{\% rms \\ IPTA}} & \multirow{2}{*}{\shortstack[l]{$T$ \\ (yr)}} &  \multirow{2}{*}{\shortstack[l]{$\Delta T$ (s) \\ $\times 10^5$ }} & \multirow{2}{*}{\shortstack[l]{$T$ \\ (yr)}} &  \multirow{2}{*}{\shortstack[l]{$\Delta T$ (s) \\ $\times 10^5$ }} & \multirow{2}{*}{\shortstack[l]{$T$ \\ (yr)}} &  \multirow{2}{*}{\shortstack[l]{$\Delta T$ (s) \\ $\times 10^5$ }} \\
		& & & & & & & \\
		\hline 
		& & & & & & & \\[-7pt]
		
		7 & 100 & 12.8 & 2.12 & 10.7 & 2.03 & 12.2 & 2.76 \\
		10 & 80 & 21.3 & 2.71 & 16.0 & 2.33 & 12.2 & 2.11 \\
		12 & 80 & 29.8 & 2.64 & 26.7 & 2.69 & 18.4 & 2.20 \\
		15 & 70 & 34.0 & 2.52 & 32.0 & 2.70 & 24.5 & 2.54 \\
		\hline
	\end{tabular}
	\caption{Adjustments made to the simulated IPTA-like array in order to fix $S/N$ of the three injected sources A, B and C. The pulsar locations are kept the same as in IPTA DR1 \citep{2016MNRAS.458.1267V}, as are the relative white noise levels of each pulsar. For $S/N \geq 10$, the noise is decreased by a constant factor in all pulsars. The cadence $\Delta T$ and observation time $T$ are averaged over for all pulsars. Then $T$ is adjusted to set the $S/N$ at specific values, keeping $\Delta T$ as close to the IPTA DR1 value as the \citet{Goldstein:2017qub} method allows.}
	\label{tab:SNRadjustments}
\end{table}

The likelihood is evaluated on a 3-D grid in amplitude ($A$) and sky location ($\theta, \phi$). $A$ is evenly sampled in log space, assuming a log flat prior between $10^{-17}$--$10^{-14}$. The location parameters $\theta$ (polar coordinate from $0$--$\pi$) and $\phi$ (azimuthal coordinate from $0$--$2\pi$) are sampled over using a grid of equal area pixels. This grid is constructed with the HEALpix algorithm \citep{HEALPix} via healpy\footnote{healpy.readthedocs.io}. HEALpix allows the user to define a grid refinement parameter $n$, which results in a number of pixels $N_{\rm pix}=12 n^2$. We choose $n = 32$, giving $N_{\rm pix}=12288$ pixels of approximately equal area of $3.36$ deg$^2$. For the likelihood calculation we use $\theta$ and $\phi$ at the middle point of each pixel.

The sky error-box $\loc$ is determined as the (smallest) area in the sky containing $90\%$ of the total likelihood. For its practical computation, the likelihood is first marginalized over A, which gives ${\cal L}(\theta,\phi)$ at each sky location. Pixels are then ranked in an array $j=1, ..., N_{\rm pix}$ in order of decreasing likelihood and their cumulative likelihood is calculated. $\loc$ is then composed by the first $K$ pixels (i.e. $j=1, ...,K$) enclosing 90\% of the total likelihood. For the sky area containing $\loc$, 
we implement the next level of HEALpix grid refinement ($n = 64$) which results in a smoother likelihood, evaluated on smaller pixels of 0.84 deg$^2$.

\subsection{Mock galaxy catalog for host selection}
Having determined ${\cal L}(A,\theta,\phi)$ we need to draw a set of properties of potential hosts from a realistic galaxy population. To this purpose, we use a mock realization of the observed sky extracted from the Millennium Run \citep{2005Natur.435..629S}. The simulation evolves dark matter particles over a volume $(500/h$\,Mpc$)^3$, reconstructing the clustering of dark matter halos. Semi-analytic galaxy formation models are then used to populate halos with galaxies, tracking their star formation, accretion and merger history. 

Although not 'state of the art', the large volume of the Millenium Run \citep[683.7 Mpc side][]{2005Natur.435..629S}, compared to more recent large scale, fully hydro-dynamical, simulations such as Illustris \citep[105.6 Mpc side][]{2014Natur.509..177V} and EAGLE \citep[100 Mpc side][]{2016A&C....15...72M}, is relevant for our work. It ensures more statistical variation in the resulting galaxies, and in particular a better sampling of the high mass tail of the distribution, which is where the best candidate galaxies reside.
We use the simulated sky maps constructed by \citet{2012MNRAS.421.2904H} that employ the semi-analytic model of \citet{2011MNRAS.413..101G}, which has been shown to reproduce a number of observed properties of galaxies, including luminosity function, morphology and clustering.

The sky maps are flux-limited to $i<21.0$ \citep[see][for full details]{2012MNRAS.421.2904H}. This results in galaxy catalogs that are complete down to stellar masses of $\approx 10^{11}\msun$ at $z=0.5$ and $\approx 4\times 10^{11}\msun$ at $z=1$. We will show in Section \ref{sec:res} that all credible hosts are above  these completeness limits. We downloaded all galaxies with stellar masses of $5\times10^{10}\msun$ and higher at $z\leq1$, which resulted in about 50 million objects. For each galaxy we store the bulge mass $\mbulge$, coordinates in the sky ($\theta,\phi$) and apparent redshift $z$. The latter is then converted to $\dlum$ by assuming our fiducial cosmology (flat $\mathrm{\Lambda CDM}$ with $h_\mathrm{0}=0.73$, $\Omega_\mathrm{M}=0.25$). This information, together with a prior on the MBHB mass ratio $q$ and the aforementioned assumptions for the $\mmbulge$-relation, is all we need to perform the calculation outlined in Section \ref{sec:Bayes}.

To limit data size, only galaxies that fall within $\loc$ are considered, which contain most of the relevant information. The simplifying assumption is made that one of the galaxies in $\loc$ is the true source of the PTA signal, but there is a $10\%$ probability it falls outside the error-box. For each galaxy, the likelihood of being the GW source host is finally computed via equation \eqref{eq:pgal}, where $A$ is determined by the injected sources and all relevant galaxy parameters are given by the mock catalogs and have prior distributions as described in Section \ref{sec:Bayes}.

\section{Results and Discussion} \label{sec:res}

For each experimental setup (injected source and $S/N$ with three random noise realizations as in Section \ref{sec:setup}), we use the null stream pipeline to obtain ${\cal L}(A,\theta,\phi)$ and determine $\loc$, the results of which we discuss here first in Section \ref{sec:loc_results}. Then, we perform the calculation as described in Section \ref{sec:Bayes} for each galaxy in $\loc$. This produces a population of $p(d|G_i)$, from which we can obtain a cumulative likelihood distribution. These results are shown in Section \ref{sec:hosts_results}.

\subsection{sky-localization} \label{sec:loc_results}
\begin{figure}
	\includegraphics[width=\columnwidth]{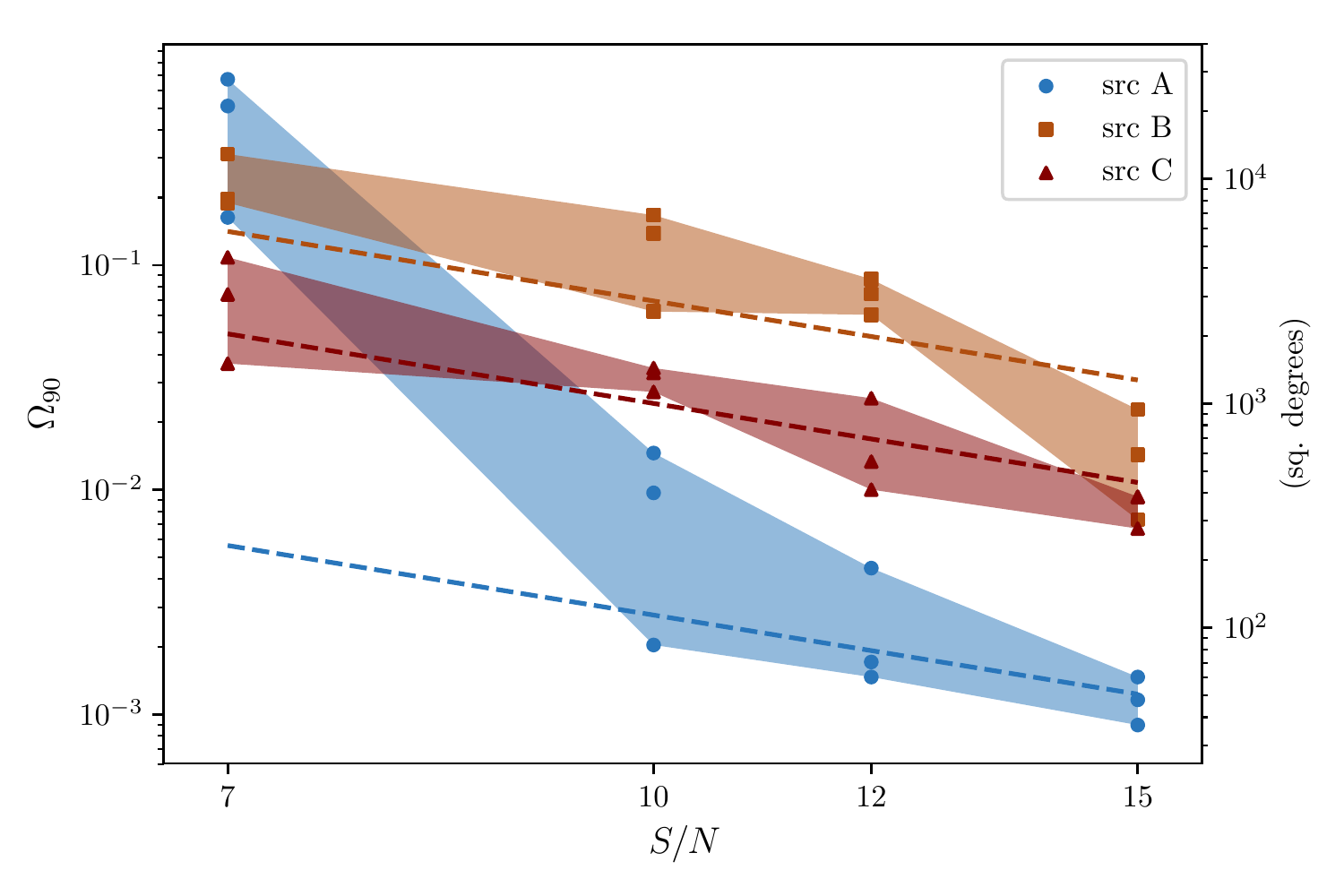}
	\caption{sky-localization accuracy for the chosen sources A, B, C at signal to noise ratios 7, 10, 12 and 15. At each $S/N$, a marker indicates $\loc$ for each of three runs with different noise realizations. The dashed lines give the best fit of $\loc \propto (S/N)^{-2}$ for the points at $S/N \geq 10$.}
	\label{fig:SNRvA90}
\end{figure}

First we look at the behavior of $\loc$ with increasing $S/N$ for the three different sources, which is shown in Figure \ref{fig:SNRvA90}. The expected trend $\loc\propto$S/N$^{-2}$ is roughly followed by all sources, albeit not perfectly, due to the small numbers of performed simulations for each case.
An exception is source A at $S/N<10$, which shows a much steeper slope. Although this is consistent with the `transition zone' identified in \citet{Goldstein:2017qub} -- signaling the $S/N$ at which the data start to be informative -- sources B and C do not behave the same way. 

\begin{figure}
	\includegraphics[width=\columnwidth]{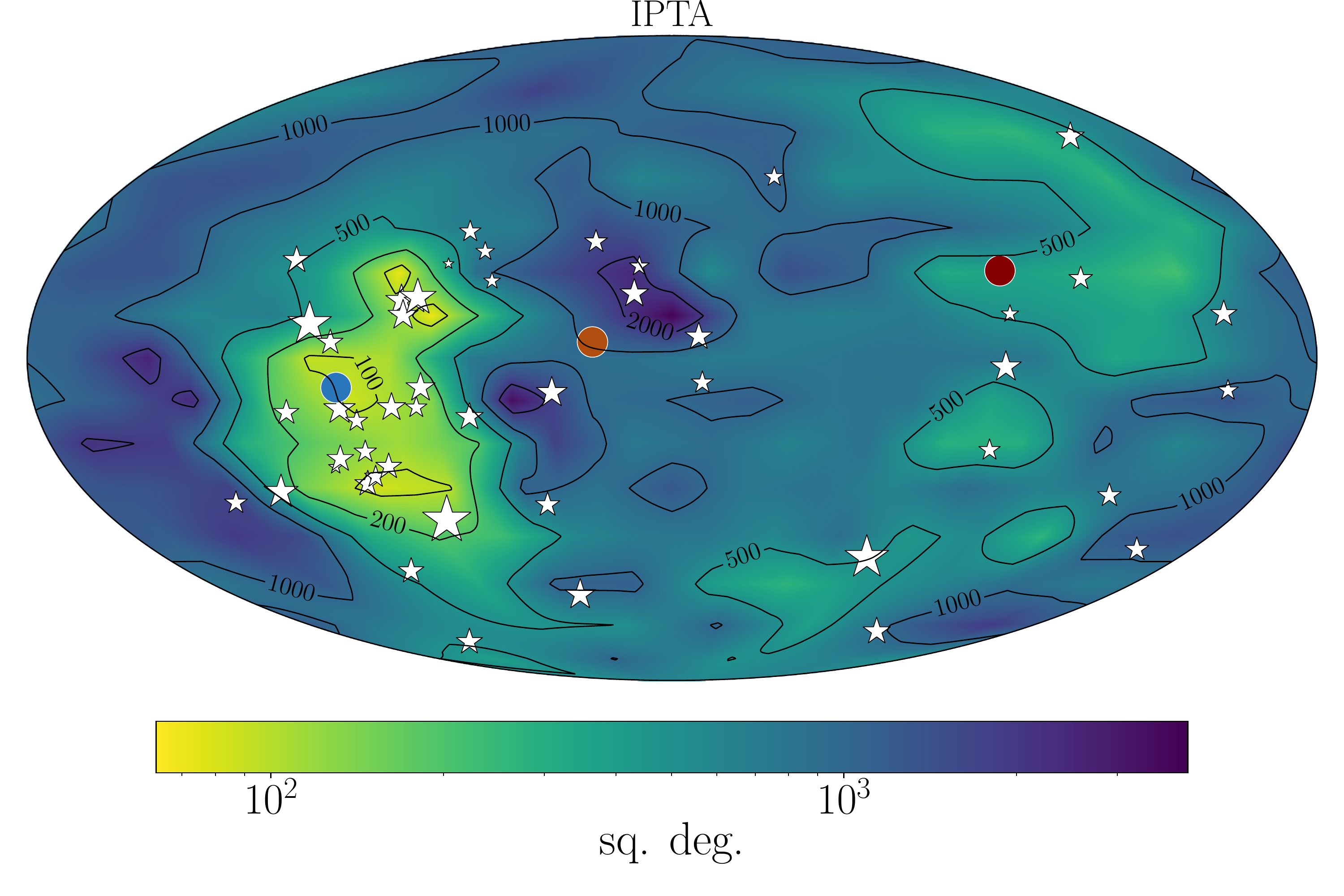}
	\caption{Localization capability of an IPTA-like array of pulsars for a source at fixed $S/N = 12$. This map is interpolated from 192 localization values obtained by injecting a source at 192 locations forming a grid of equal sky area pixels (a HEALPix grid with $n=4$ \citep{HEALPix}). The IPTA pulsars are marked with stars, where the size of the star corresponds to the noise level of the pulsar (with bigger stars for lower noise). The circles indicate the positions of sources A (blue, left), B (orange, middle) and C (red, right).}
	\label{fig:localization_map}
\end{figure}

We conjecture that this is related to the specific position of the sources, relative to the pulsars (see Figure \ref{fig:localization_map}). 
When the source is close to the location of the best pulsars (like A), the combined S/N from all pulsars at the marginal detection level (S/N$\approx7$) is mostly due to the contribution of these few, good pulsars (or possibly only one good pulsar). The other pulsars have a very low individual S/N. Therefore, the source is effectively triangulated by very few pulsars, making localization poor.
At higher total S/N$\approx$10, more pulsars contribute to the triangulation as their individual S/N increases. As such, their is a steep improvement in sky-localization, steeper than the canonical (S/N)$^{-2}$ slope.

Conversely, when the source is far away from the majority of the best pulsars (like B and C), a detection with S/N$\approx 7$ already requires contribution from several different pulsars, making triangulation more effective. After this transition (the shaded area crossing in Figure \ref{fig:SNRvA90}), the standard $S/N$ scaling continues for source A as well.
\\

Apart from the trend, the localization accuracy of the three sources vary by a factor of $\sim\! 20$ between them. 
This is due to both the inhomogeneous distribution of pulsars in the sky \citep{2010PhRvD..81j4008S} as well as the different quality of the pulsars in the arrays \citep{doi:10.1093/mnras/stv2092}, which is expected to cause a difference in localization.
The best localization, at high $S/N$, is achieved for source A, sitting in the `sweet spot' of the array (where most of the pulsars, including the best ones, are). However, there is not simply a monotonic increase of $\loc$ for sources further away, since the furthest source C has a better localization than source B.
This is also expected since, due to the shape of the PTA response function, sources that are antipodal to the sky region that is best covered by the array are better localized than sources that are orthogonal to that region \cite[see, e.g., figure 10 in][]{2010PhRvD..81j4008S}

A further investigation of this is visualized in Figure \ref{fig:localization_map}. Here we inject a source with the same parameters as A at 192 different locations in the sky into white noise, using a synthetic IPTA-like array. The $S/N$ is set to 12 everywhere, by scaling the amplitude $A$ of the GW signal. The map shows the resulting localization $\loc$ at each point.
A dipolar structure of $\loc$ is noticeable, where sources near the `sweet spot' of clustered pulsars  -- which includes most of the best pulsars -- and to a lesser extent, sources near the antipodal point are localized better than sources in between. This is related to the quadrupolar nature of GWs, which results in a pulsar response function that has this antipodal symmetry, as was also shown by \citet{2010PhRvD..81j4008S}.

In any case, the huge scatter in $\loc$ warns of a potential risk of an anisotropic sky coverage of the pulsars in the array. Should the loudest resolvable GW sources be positioned at unfavorable locations, their detection, even at moderate S/N $\approx12$, would allow sky-localization accuracies of about \mbox{$2000$ deg$^2$} only (an area containing $\sim$2 million galaxies in our catalog before any selection), jeopardizing any effort to identify a possible EM counterpart.

\subsection{Host candidate population} \label{sec:hosts_results}

\subsubsection{Number of credible host candidates}
Our main results consist of a set of $p(d|G_i)$ for the galaxies $\{G_i\}$ within $\loc$ for each experimental setup (Section \ref{sec:setup}). First, we compute the cumulative likelihood distribution from these $p(d|G_i)$. We then define $N_\mathrm{x}$ to be the minimum number of galaxies needed to sum to x\% of the total likelihood $\sum_i p(d|G_i)$. Specifically, we look at $\nfifty$ and $\nninety$ as proxies for the expected number of candidate host galaxies.

\begin{figure}
	\includegraphics[width=\columnwidth]{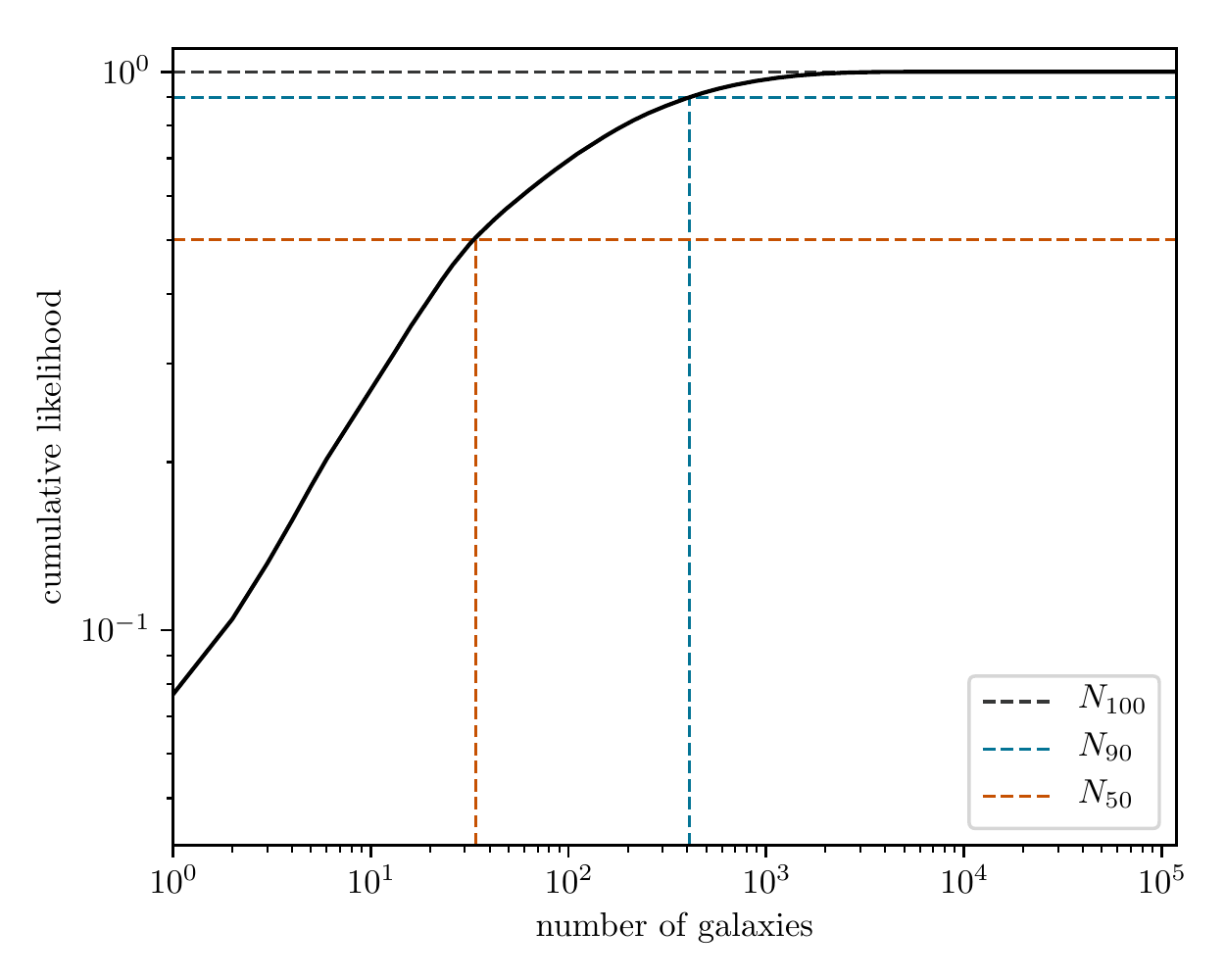}
	\caption{Cumulative likelihood of $p(d|G_i)$. The likelihood data $d$ is for the IPTA setup (as described in the text) with source A at $S/N=15$ (one of the random noise realizations). Vertical dashed lines identify the number of galaxies making up 50\% (orange) and 90\% (blue) of the total likelihood.}
	\label{fig:cumu_like}
\end{figure}

An example can be seen in Figure \ref{fig:cumu_like} for source A at $S/N = 15$ (the first random noise realization). Within $\loc\approx 60$ $\deg^2$, there are ${\sim}1.2\times10^5$ galaxies $\{G_i\}$ in our mock catalog, which would make detailed follow-ups for host identification impractical. The potential benefit of our technique is apparent from the fact that of those galaxies, only $\nninety = 409$ make up $90\%$ of $p(d|G_i)$, and $\nfifty = 34$ make up $50\%$ of $p(d|G_i)$. 
\\

\begin{figure}
	\includegraphics[width=\columnwidth]{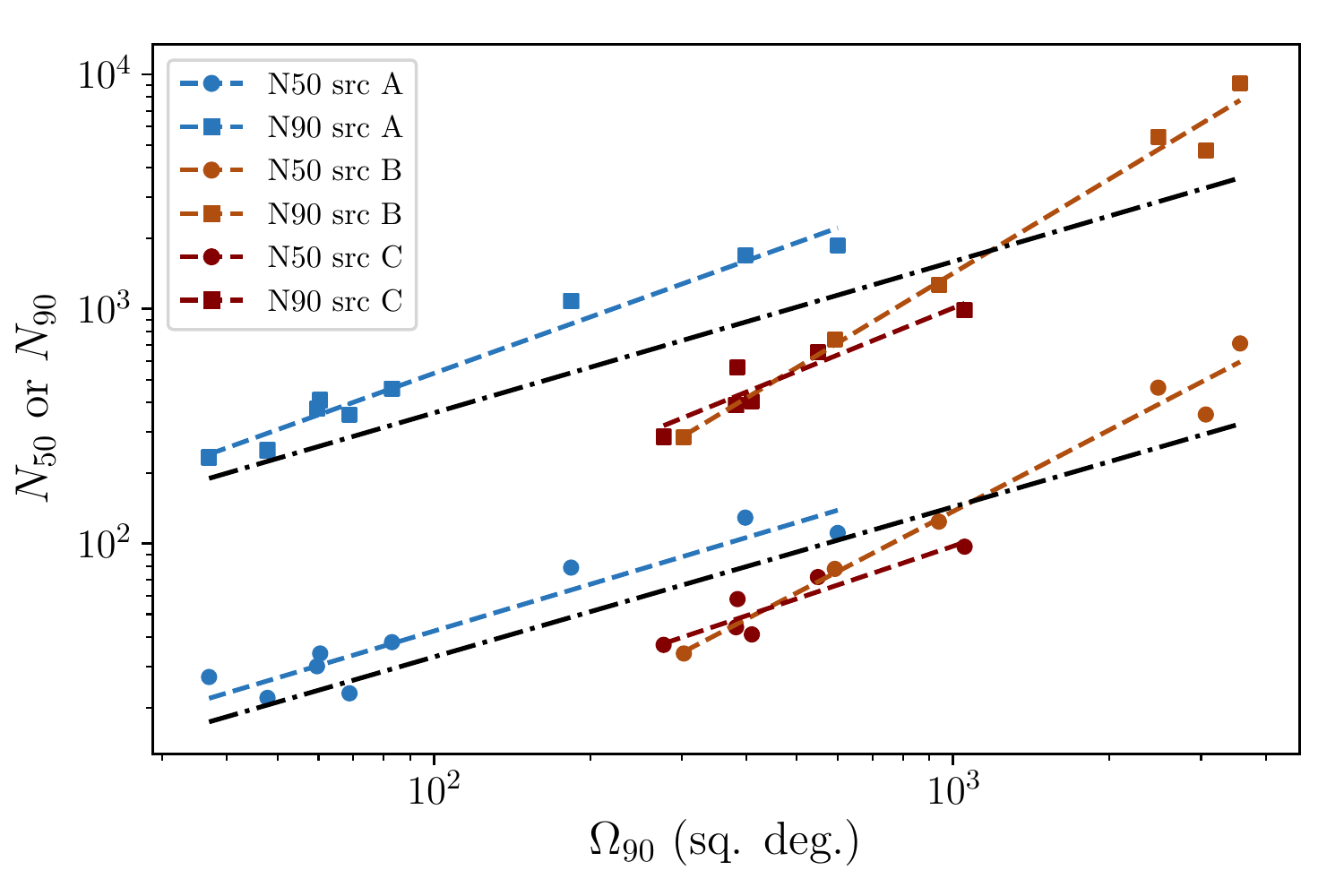}
	\caption{Number of candidate galaxies adding up to $50\%$ ($\nfifty$, circle markers) or $90\%$ ($\nninety$, square markers) of the total likelihood to host the detected source, versus the sky-localization accuracy $\loc$ for that detection. Results are shown for source A (blue) at $S/N =$ 10, 12 and 15, and sources B (orange) and C (red) at $S/N =$ 12 and 15. For each $S/N$ three noise realizations give a cluster of points at similar $\loc$ values. Dashed lines show fitted power laws per source (see Table \ref{tab:fit_powers} for the best-fit powers). Dot-dashed lines are fits to $\nfifty$ for all sources, with power 0.64, and to $\nninety$, with power 0.65.}
	\label{fig:Nx_A90}
\end{figure}

The collection of $\nfifty$ and $\nninety$ of all experimental cases for which we obtained results can be found in Figure \ref{fig:Nx_A90}. We can fit a power-law as $N_\mathrm{x} = c (\loc/\loc^*)^p$, with parameters $c$, $p$ (the power) and $\loc^*$, and $\mathrm{x}$ being either 50 or 90. By minimizing the sum of squared differences between the predicted log values and the log of the data points, we obtain best fitting powers 0.64 and 0.65, for $\nfifty$ and $\nninety$ respectively. Although naively one would expect a linear proportionality between $\loc$ and the number of potential hosts, there is a significant scatter on the relation.

Tighter fits are obtained by treating the points for different injected sources separately, with best fit powers as in Table \ref{tab:fit_powers}. These numbers show that fits to individual source data points are generally steeper and closer to the expected linear dependence. One of the causes of the shallower global fit appears to be the larger $\nfifty$ and $\nninety$ for source A with respect to sources B and C at sky-localizations of $\approx 300$\,deg$^2$, as shown in \ref{fig:Nx_A90}. (Source A has $S/N = 10$ around this localization accuracy, while source B and C have higher $S/N = 15$. Consequently, $\nninety$ and $\nfifty$ for source A includes galaxies with a lower bulge mass than for B and C, resulting in a larger $\nninety$ and $\nfifty$).

So while there is clearly a relation between the size of the sky error-box and the number of candidate host galaxies, scatter is caused by factors related to the detailed source properties. Nonetheless, as a rule of thumb, we expect that for a resolvable PTA signal located in the sky with a precision of $\approx 100$\,deg$^2$, we can identify few hundreds (few tens) galaxies in which the source sits with 90\% (50\%) confidence. Compared to all galaxies with stellar mass $>5\times 10^{11}\msun$ at $z<1$ falling in the error-box, these numbers restrict the pool of realistic hosts by nearly three (four) orders of magnitude, making realistic detailed follow-up campaigns feasible.

\begin{table}
	\centering
	\begin{tabular}{ccc}
		\hline
		& & \\[-7 pt]
		source & $\nfifty$ power & $\nninety$ power \\
		\hline
		A & 0.66 & 0.80 \\
		B & 1.15 & 1.34 \\
		C & 0.74 & 0.90 \\
		all & 0.64 & 0.65 \\
		\hline
	\end{tabular}
	\caption{Best fit powers for the power law fits to $\nfifty$ and $\nninety$ as in Figure \ref{fig:Nx_A90}. These are obtained by minimizing the sum of squared errors on the log $N_\mathrm{x}$ values.}
	\label{tab:fit_powers}
\end{table}

We also calculated $p(d|G_i)$ and $N_\mathrm{x}$ for source A at $S/N = 7$, which has a very poor sky-localization of about $2.8 \times 10^4$ sq. degrees (67\% of the sky). The expected number of candidate hosts becomes very large, and also disobeys the trend discussed above. We conjecture this is due to the localization likelihood distribution not having a single peak for the low $S/N$ case, so potential hosts are allowed to be anywhere in the localization error-box, which is most of the sky.\\

\subsubsection{Host candidate sky distribution and clustering}
\begin{figure}
	\includegraphics[width=\columnwidth]{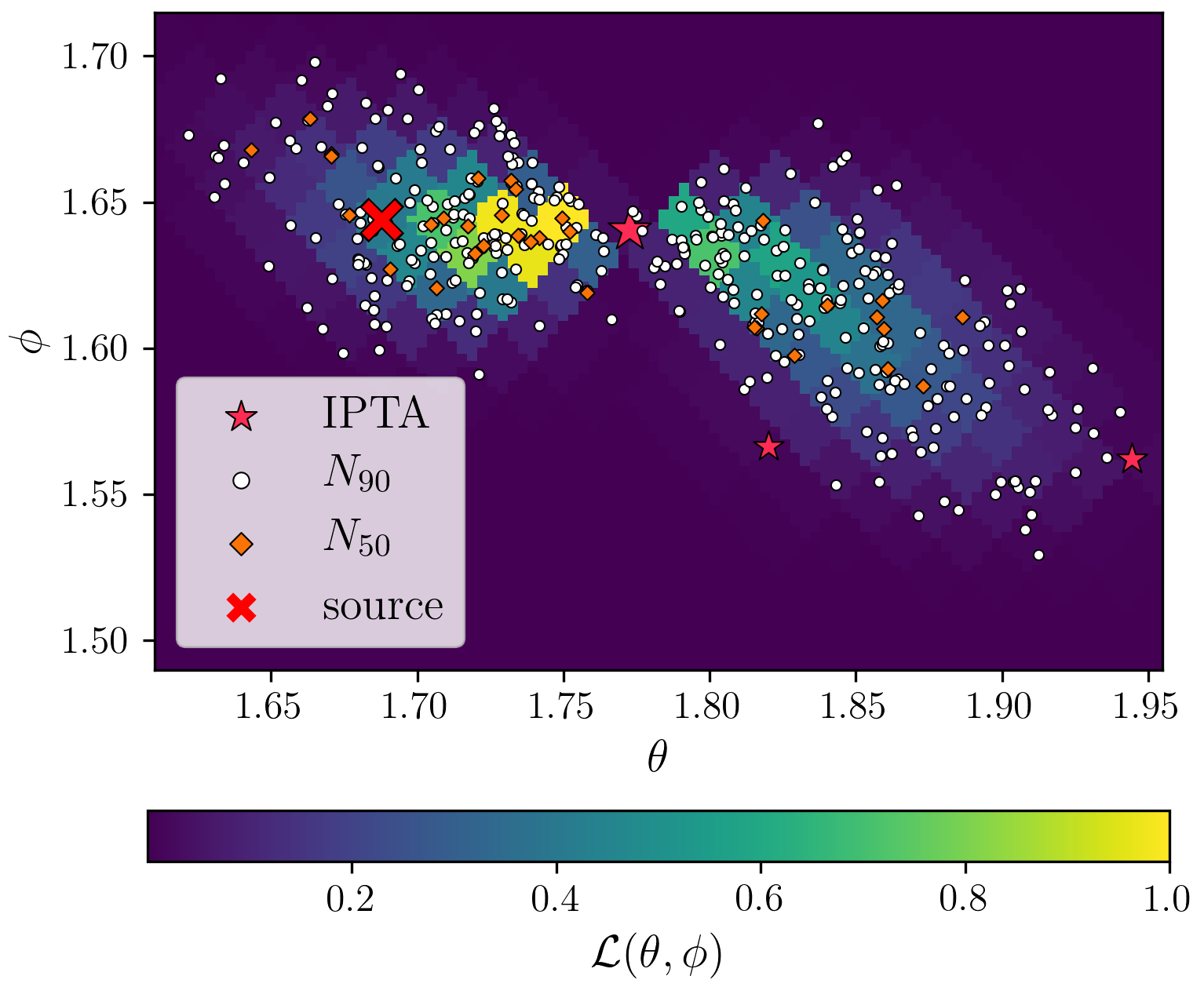}
	\caption{Locations of the best candidate host galaxies on top of the sky location likelihood for the injected source A (located at the red cross). The PTA has pulsar locations (pink stars) and relative noise levels of the IPTA DR1, but is adjusted such that the total $S/N = 15.0$ (see text). The 34 best candidates sum to $50\%$ of the likelihood to be the host galaxy ($\nfifty$ in orange diamonds) and an additional 375 sum to $90\%$ ($\nninety$ in white circles). For this example, the $M_{BH}-M_{bulge}$-relation is marginalized over priors obtained from the literature (see text).}
	\label{fig:galaxy_locations}
\end{figure}

Apart from the number of galaxies that make up a significant fraction of the likelihood $\sum_i p(d|G_i)$, we can also look at the properties of these galaxies. The parameters from the mock galaxy catalog are $\mbulge, \dlum,\theta, \phi$. First, the sky locations of galaxies within $\nfifty$ or $\nninety$ for the example case (source A at $S/N = 15$), are shown in Figure \ref{fig:galaxy_locations}. They are plotted on top of the localization likelihood $\mathcal{L}(\theta, \phi)$ of the injected source. 
The galaxies follow the shape of the localization area, because we only used galaxies within $\loc$. Moreover, it can be seen that there is a relatively higher concentration of $\nfifty$ galaxies in the highest likelihood pixels. Hence, $\mathcal{L}(\theta, \phi)$ must contribute more to the selection of candidates than simply what we get from selecting the ones in $\loc$.

\begin{figure}
	\includegraphics[width=\columnwidth]{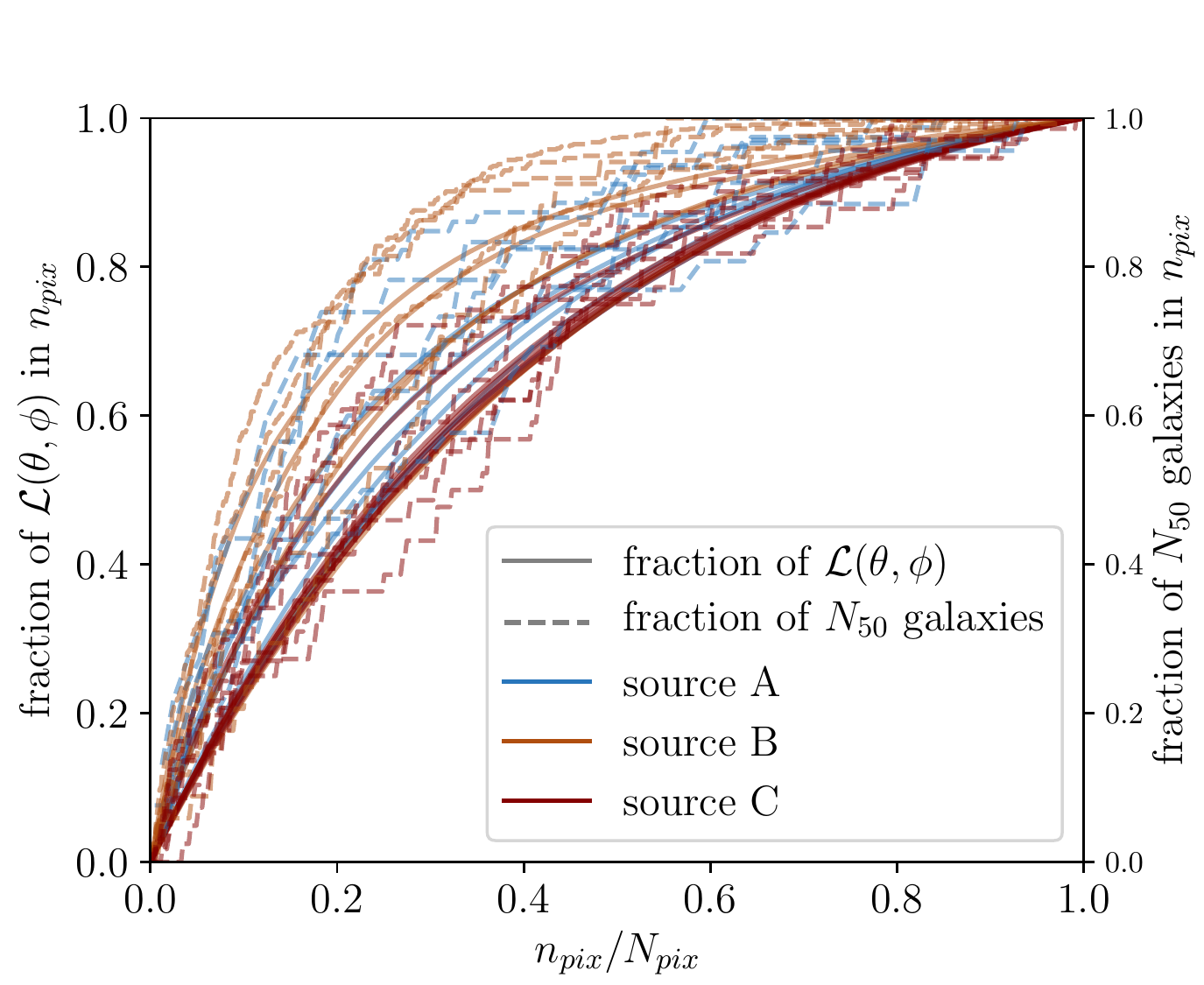}
	\caption{Comparison between the concentration of the sky-localization likelihood and of the locations of good candidate host galaxies. With the fractional area of $\loc$ on the x-axis, the fractional localization likelihood in this area on the first y-axis (left, solid lines), and the fractional number of $\nfifty$ galaxies on the second y-axis (right, dashed lines) (see text for details.) The quantities are normalized between 0 and 1 so that all experimental cases fit on the same scale. This plot includes all three injected sources (A in blue, B in orange and C in red), for $S/N = 12,\, 15$.}
	\label{fig:galaxy_clustering}
\end{figure}

We further investigate this statement using the clustering of good candidate galaxies -- the $\nfifty$ galaxies -- for all of the experimental cases. Figure \ref{fig:galaxy_clustering} simultaneously shows a measure of the concentration of the localization likelihood $\mathcal{L}(\theta, \phi)$, and of the concentration of $\nfifty$ galaxies. The sky error-box $\loc$ consists of a number of pixels $N_\mathrm{pix}$ that are sorted in descending $\mathcal{L}(\mathrm{pix}_i)$ order. Starting with the best pixel, we iteratively increase this number by adding the next best pixel. The size of the included area is recorded as the fraction of the number of pixels over the total in $\loc$, i.e. $n_\mathrm{pix} / N_\mathrm{pix}$. The concentration of the localization likelihood then is the likelihood in $n_\mathrm{pix}$ as a fraction of the total, i.e. $\sum_i^n \mathcal{L}(\mathrm{pix}_i) / \sum_i^N \mathcal{L}(\mathrm{pix}_i)$. 

We compare this with the concentration of good candidate hosts, as the fractional number of $\nfifty$ galaxies in the selected pixels. The distribution are spread out, but there is no significant difference between the sky likelihood and candidate host concentration, i.e. the host probability follows the sky-localization distribution. We therefore conclude that it is valuable to include detailed sky-localization information when selecting candidate host galaxies, rather than only making a selection based on the total sky-localization area.\\

\subsubsection{Host candidate mass and redshift} \label{sec:mass_redshift}
\begin{figure}
	\includegraphics[width=\columnwidth]{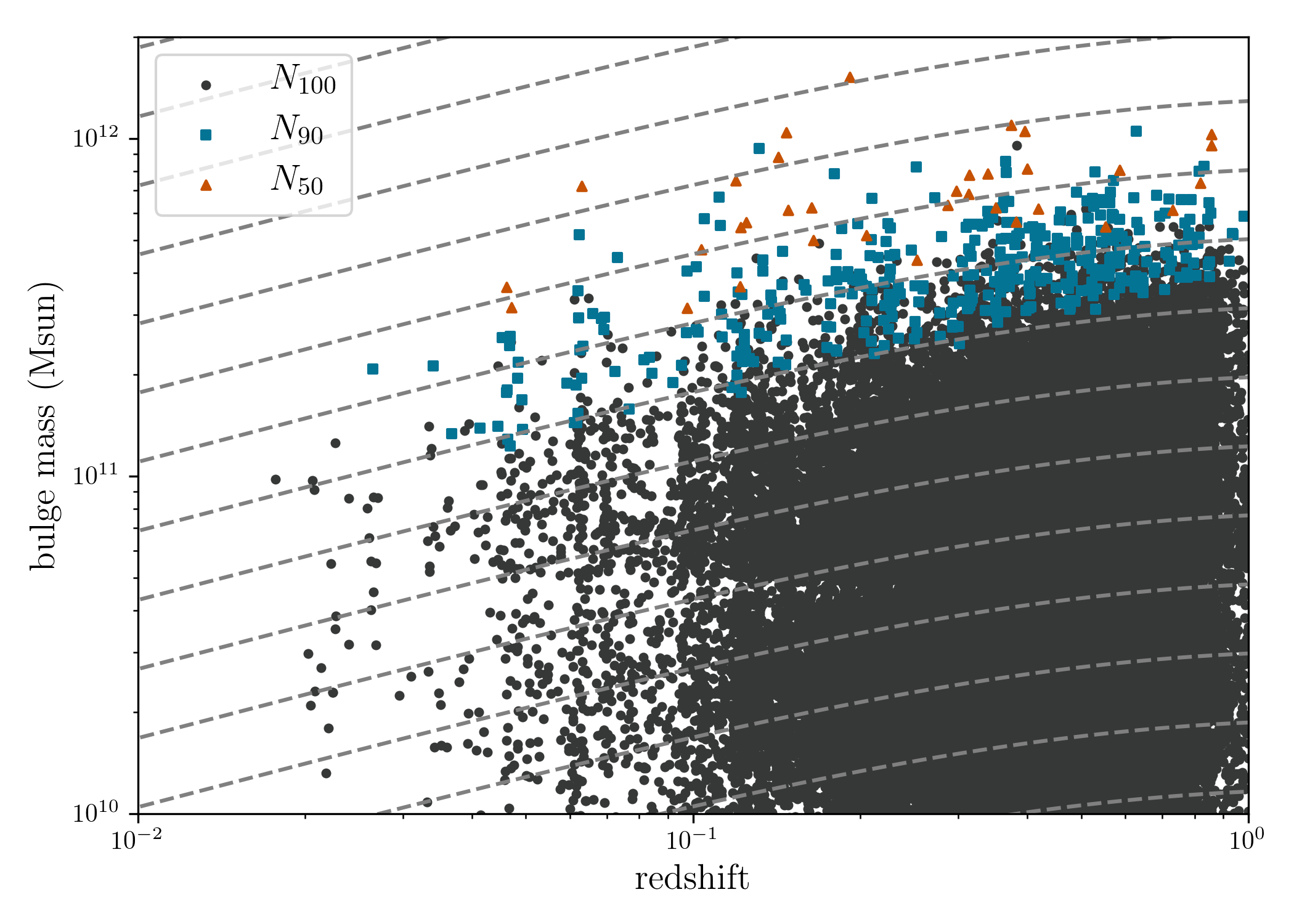}
	\caption{Distribution of bulge masses and redshifts of the candidate host galaxies of the example case source A with $S/N$ = 15. Blue squares mark galaxies that make up $\nninety$ and orange triangles mark the best candidates, which make up $\nfifty$. All other galaxies, that fall within the sky-localization error-box $\loc$, but form the lowest 10\% of the total likelihood $\sum_i p(d|G_i)$, are marked with (dark) gray circles. Dashed gray lines are lines of constant GW amplitude (as in Equation \ref{eq:A}).}
	\label{fig:Mbulge_redshift}
\end{figure}

Second, we consider the other two parameters from the catalog, the bulge mass $\mbulge$ and luminosity distance $\dlum$. Figure \ref{fig:Mbulge_redshift} shows their distribution among candidate hosts for the example case, where $\dlum$ has been converted into redshift. This figure best visualizes the key idea behind our method. Since $A\propto M^{5/3}/\dlum$ -- and there is a proportionality $M\propto\mbulge^{\beta}$ and an almost linear proportionality between $\dlum$ and $z$ at $z<1$ -- there is only a stripe in the mass--redshift plane defining the region of possible galaxy hosts. Moreover, since the first detection of a resolved PTA source will necessarily involve a very strong signal from a very massive binary system, this region lies at the highest masses. Due to the steep decay of the high mass end of the galaxy mass function, only few credible host candidates can be identified. 

In the example shown, galaxies belonging to $\nfifty$ or $\nninety$ are bound by a line of slope $3/(5\beta)$ in the $\log{\dlum(z)}-\log{\mbulge}$ plane (where $\beta$ is the $\mmbulge$ constant marginalized over our prior), as expected by the GW amplitude scaling. There is however a large mixing of galaxies with different likelihoods in this plane, due to their specific sky location. For example, there are a few very massive galaxies that fall into the lowest 10\% of the likelihood, which is due to an unfavored sky position. Note that there are $\nfifty$ candidates across the whole range of redshifts in our sample.

\begin{figure}
	\includegraphics[width=\columnwidth]{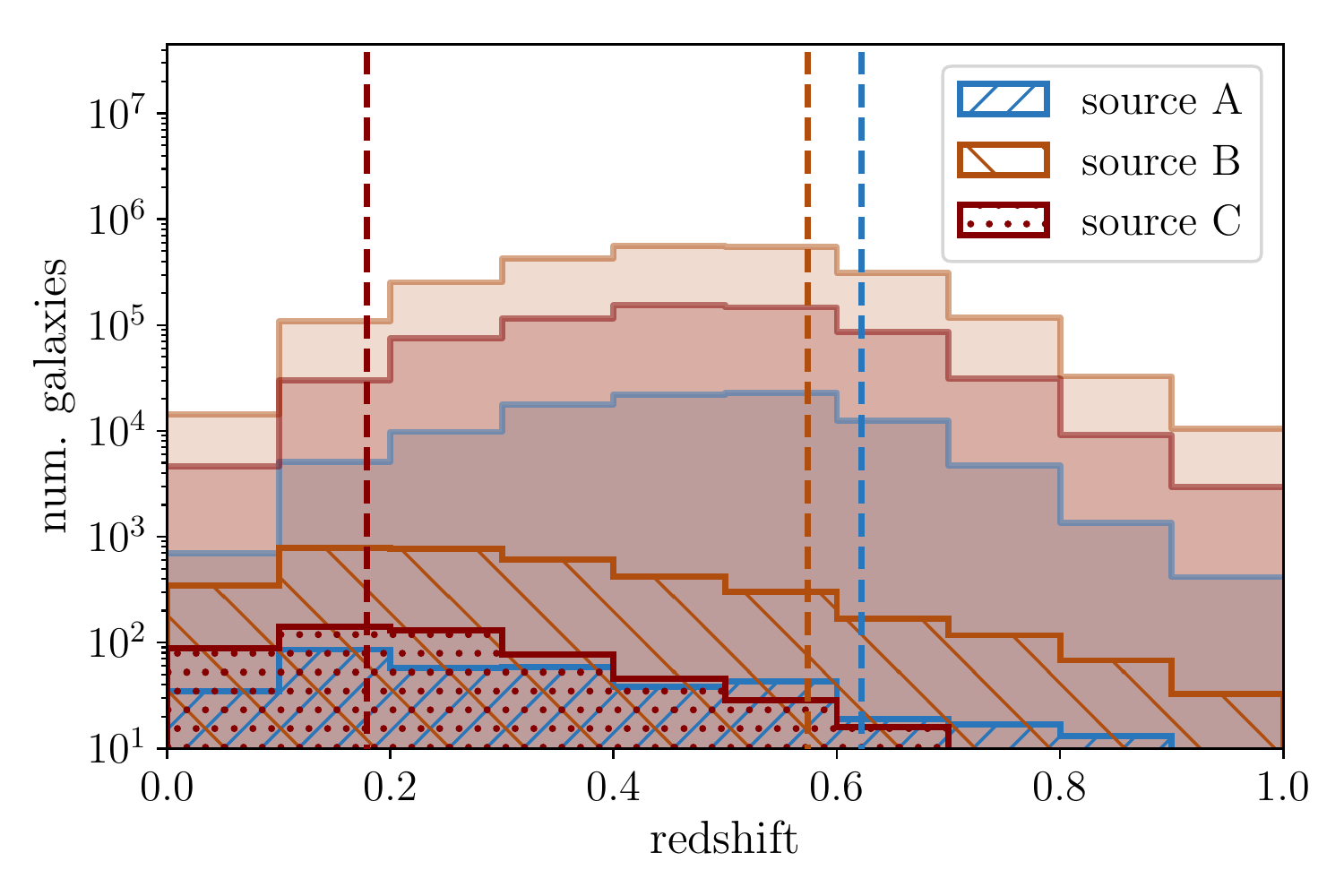}
	\caption{Logarithmic histogram of the redshifts of candidate host galaxies per source. The counts from the six experimental cases with $S/N = 12, 15$ and three noise realizations are averaged over. The foreground (hatched) histograms are $\nninety$ candidates, and the background (filled) histograms are all (i.e. $N_\mathrm{100}$) candidates from the selected sky error-box. Injected redshift values for each source are indicated by a dashed line (see also Table \ref{tab:source_params}).}
	\label{fig:redshifts}
\end{figure}

The redshift of the injected source A is 0.62 (Table \ref{tab:source_params}), so it's not a surprise that candidate hosts for this source have redshifts across the whole range 0--1. To explore this further we look at the redshifts of candidate host galaxies for all injected sources and $S/N$ values. Figure \ref{fig:redshifts} shows a number of histograms of $z$ on a logarithmic scale. For each source A, B and C, the results of $S/N = $12 and 15 (with three noise realizations per $S/N$), are combined. We make a comparison between the redshift distribution of the candidate galaxies pre-selected within the sky error-box (the background histograms), and the $\nninety$ candidates selected with out method (the foreground, hatched histograms).

Compared to the prior distribution, lower redshifts are preferred. However, for all injected sources, there are a significant number of candidates at redshifts $z > 0.6$. Even though the injected redshift for source C ($z=0.18$) is much lower than four sources A and B, the redshift distributions of candidate hosts differ only slightly, which reflects the fact that redshift is degenerate with mass in our method.\\

The turnover in the total number of galaxies in the error-box seen in figure \ref{fig:redshifts} at $z>0.5$ is due to the $i=21$ flux limit of the adopted galaxy catalogs, which result in severe incompleteness of lower-mass galaxies. Figure \ref{fig:Mbulge_redshift}, however, shows that typical galaxies belonging to $\nninety$ have $M_b>2\times 10^{11}$ at $z=0.5$ and $M_b>4\times 10^{11}$ at $z=1$. The adopted catalog is therefore complete in the mass-redshift range where potential GW galaxy hosts live.

Figure \ref{fig:Mbulge_redshift} also shows that the distribution of credible galaxy hosts of resolvable PTA sources peaks at $z\lesssim0.2$,  whereas two out of three of our selected signals (A and B) lie around $z\approx 0.6$. GW sources were picked according to their sky location, therefore A, B, C are not an unbiased sample and are not necessarily representative of the actual redshift distribution of the first GW resolved signals. However, there are several systems at $z>0.5$ in the sample of 30 resolvable sources found in Section \ref{sec:setup}, and also \cite{2015MNRAS.451.2417R} found that the peak of the first resolved PTA sources is at $z\approx 0.5$. 

High $z$ sources are more common despite there being less potential host galaxies at such redshifts. This indicates that the likelihood of a galaxy to be a host is not only connected to its sky location and its position in the $M$--$z$ plane, considered in this work. The other key parameter is likely to be the absolute galaxy mass (regardless of redshift). There is evidence -- both from observations and from cosmological simulations \citep[see, e.g., results compiled in figure 1 of][]{2016MNRAS.463L...6S} -- that the galaxy merger rate at low redshift is a strong function of the galaxy mass, with massive galaxies merging more often. 

Since the MBHB population simulated in Section \ref{sec:setup} consistently takes this fact into account, the resulting MBHB population is naturally skewed towards high masses. Conversely, our host selection method only picks galaxies based on the GW amplitude given the combination of redshift and bulge mass, and therefore chooses relatively more lighter galaxies. However, because of these candidates' lower masses, they are less likely to have undergone a major merger (and hence host a GW source) compared to the few more massive ones picked at higher redshift. This suggests that combining our method with a (prior) probability of hosting a MBHB based on galaxy mass only \citep{2014MNRAS.439.3986R,2017NatAs...1..886M} can somewhat break the intrinsic mass--redshift degeneracy, further reducing the numbers of credible galaxy hosts.

\subsection{Assumptions and Approximations} \label{sec:disc}
Although simulations performed in this work are realistic in many aspects, few assumptions and choices had to be made to make their total runtime manageable. \\

Several assumptions were made in the connection of the chirp mass of the GW source to the bulge mass of the host galaxy. 
First, we assumed a log-flat prior on $-2\leq {\rm log}q \leq0$, based on the broad $q$ distribution of merging binaries found in cosmological simulations \citep{2017MNRAS.464.3131K}. Although this is not necessarily representative of the $q$ distribution of real MBHBs, we tested that different choices have only a minor impact on the results \citep[see also][]{2018ApJ...856...42S,2018ApJ...863L..36I,holgado_pulsar_2018}.

Second, we did not consider errors in the measurements of galaxy $\mbulge$ and $\dlum$. The latter does not matter; for any practical purposes, galaxy redshifts can be determined almost exactly, and estimates of $\dlum$ are only affected by galaxy peculiar velocities and uncertainties in the knowledge of the cosmological parameters, resulting in a negligible few\% error. Conversely, the former can be significant, as bulge mass determination can be uncertain within a factor of two. This is likely to impact our results, spreading the host probability distribution thus returning more host candidates. Some tests on a limited number of setups found that including an uncertainty of a factor of two on the galaxy bulge mass results in roughly a factor of two more candidate hosts galaxies.

Last, we marginalized over the uncertainty in the $\mmbulge$-relation. Assuming a specific $\mmbulge$-relation instead can affect our results, especially if the relations predicts relatively higher or lower black hole masses than the marginalized relation. As an example, we ran some test cases assuming the $\mmbulge$-relation from \citet{korho}, which associates relatively higher black hole masses given the galaxy bulge mass. The number of candidate host galaxies in these cases is increased by a factor ranging between $\sim\! 3$ and $\sim\! 8$ with respect to the marginalized $\mmbulge$ case. Conversely, for a `pessimistic' $\mmbulge$-relation such as \citet{shankar} -- which predicts relatively lower black hole masses especially for high-mass galaxies -- the number of candidates is a factor $\sim\! 2$ to $\sim\! 4$ lower.\\

Due to computational limitations, we ran a limited number of simulations. Although we checked robustness of the results against the specific noise realization, we only picked one sky location for each source. This may make cosmic variance a factor in the determination of the number of galaxy hosts. To test this, for a selected GW  source, we performed some rigid rotations of the Millennium sky, and counted $\nfifty$ and $\nninety$ for each of them. Although numbers vary, the scattering is consistent with that observed in figure \ref{fig:Nx_A90}.

An important assumption of our method is that the true host of the detected GW signal is present in the galaxy catalog. This is guaranteed only for complete catalogs. Real catalogs based on observations never are, and the simulated catalog from \citet{2012MNRAS.421.2904H} reflects this by selecting galaxies based on observational criteria. This results in a number of missing galaxies -- more towards higher redshifts. However, for the most part these are the small galaxies (which are more difficult to observe), and those are not relevant host candidates. Since at redshifts $z \lesssim 1$ only the most massive galaxies are selected in $\nninety$ (see Figure \ref{fig:Mbulge_redshift}), this is unlikely to affect the results for $\nninety$ and $\nfifty$, but it is a possible source of error. As there are good candidates up to $z=1$, it is also possible there are a small number of potential $\nninety$ galaxies at $z>1$ that were not included.

Finally, it should be kept in mind that we selected the 90\% sky location credible region. By selecting $\nfifty$ and $\nninety$ in this region, the actual probability to find the true host in these sets is $0.9\times 0.5=0.45$ and $0.9\times 0.9=0.81$, respectively.

\section{Conclusion} \label{sec:conc}
In this paper, we proposed a novel methodology to select host galaxy candidates of the first individual gravitational wave sources observed by pulsar timing arrays. Since PTA source localization is expected to be of several deg$^2$ at best, up to several million galaxies might end up in the sky error-box. Classifying the most promising host candidates is therefore of paramount importance to increase the chances of true host identification via dedicated follow-ups. Our method exploits the GW strength dependence on chirp mass and distance, together with empirical MBH mass--host galaxy correlation, to rank galaxies in the mass--redshift plane. We frame this concept in the Bayesian language, together with the null-stream based sky-localization method developed in \citet{Goldstein:2017qub}, to assign each galaxy a probability of hosting the MBHB generating a specific GW signal.\\

To test our method, we performed realistic simulations by drawing GW sources from detailed MBHB population models based on observed merging galaxies, by employing the actual IPTA pulsar sky locations and rms values to build the array, and by selecting host candidates based on formation and evolution models. We considered different GW source sky positions and detection S/N and investigated the ensemble of credible host galaxy candidates. In particular, we defined $\nfifty$ and $\nninety$ to be the smallest numbers of galaxies having a collective 50\% and 90\% chance of being the true host of the GW source, respectively, assuming the true host is among the prior selection of candidates. Our key results can be summarized as follows:
\begin{itemize}
\item $\nfifty$ and $\nninety$ are respectively nearly four and three orders of magnitude smaller than the number of galaxies with stellar mass $M_*>5\times10^{10}\msun$ at $z<1$ found in the 90\% confidence sky location region $\loc$;
\item $\nfifty$ and $\nninety$ should roughly be proportional to $\loc$. We find a sub-linear proportionality, although with large scatter;
\item despite the large scatter, a useful rule of thumb is that for $\loc=100$deg$^2$, $\nfifty\lesssim 50$ and $\nninety\lesssim 500$;
\item although the distribution of potential hosts peaks around $z<0.2$, it has a long tail that extends up to $z\lesssim 1$. 
\end{itemize}

Our methodology can therefore effectively select the most likely host galaxy candidates, which might have a major impact on future multi-messenger observations of MBHBs. For typical PTA sky-localization precision of hundreds of deg$^2$, instead of following up millions of galaxies, we can choose to accept the risk of missing the true host with 55\% (19\%) probability and monitor only the $\approx 100$ (1000) most promising ones. There is significant uncertainty on these numbers, mainly due to the uncertainty in the $\mmbulge$-relation (see Section \ref{sec:disc}).\\

The applicability of our method obviously relies on the availability of photometric and spectroscopic data from all-sky surveys necessary to identify potential galaxy hosts and to estimate their stellar (and bulge) masses. Since the most credible galaxy candidates are necessarily very massive (and/or particularly nearby), relatively shallow surveys are sufficient tfor this scope. Catalogs from SDSS \citep[][covering $\approx1/4$ of the sky]{2015ApJS..219...12A}, Pan-STARRS \citep[][$\approx3/4$ of the sky]{2002SPIE.4836..154K}, LSST \citep[][$\approx1/2$ of the sky]{2009arXiv0912.0201L} and Gaia \citep[][all sky]{2016A&A...595A...1G} will provide enough imaging, photometric and (possibly) spectroscopic information for reliable mass estimates via, e.g. spectral energy distribution fitting \citep[see e.g.][]{2009MNRAS.394..774L,2014MNRAS.444.2960D}.

Note that a positive host identification chance increase of less than a factor of two comes at the expense of following up a factor of ten more galaxies. The follow up strategy can therefore be optimized based on the future number of resolved PTA sources and on available observing facilities. Reducing the number of credible host is critical mostly because our knowledge of MBHB signatures is poor \citep[see, e.g.][]{2012AdAst2012E...3D}. One therefore has to collect all possible hints to build up confidence that the true host have been found. This might require, for example, multiple photometric and spectroscopic optical and IR follow up of the candidates to unveil any observational hint of an accreting MBHB, deep field imaging to assess the presence of post merger features such as stellar tails and shells \citep[e.g.][]{2008MNRAS.391.1137L}, integral field spectroscopy to identify the presence of a `dry' MBHB via kinematic signatures in the stellar distribution \citep{2013MNRAS.433.2502M}, deep X-ray observations to unveil the presence of an obscured AGN and it's possible high energy signatures \citep{2018Natur.563..214K}, and many more. 

The upcoming ELT \citep{2007Msngr.127...11G} and JWST \citep{2006SSRv..123..485G} will be particularly suited for the optical and near infrared follow-ups mentioned above, whereas the X-ray satellite Athena \citep{2013arXiv1306.2307N} can potentially survey the 100 most probable hosts within less than 1 day of observation time. Clearly, the fewer the candidates, the more extensive the follow-up campaign can be, thus enhancing the chances of a positive detection. Archival data can also be used to identify hints of, e.g, periodic variability matching the frequency of the GW source. This can be done in the optical and, possibly, in X-ray with LSST and eROSITA \citep{2012arXiv1209.3114M} archival data respectively.\\

Finally, the mismatch between the credible host redshift distribution identified with our method and the expected distribution of the first PTA sources predicted by \cite{2015MNRAS.451.2417R} indicates that a more efficient galaxy host selection can be performed when the mass-dependent galaxy merger probability is folded into the calculation \citep[see also][]{2017NatAs...1..886M}. By doing so, the mass--redshift degeneracy intrinsic in our method might be alleviated, further decreasing the number of credible hosts. We plan to further pursue this line of investigation in future work.

\section*{Acknowledgements}
We thank A. Vecchio for useful comments. A.S. is supported by the Royal Society. J.~V. is supported by STFC grant ST/K005014/1. A.M.H. is supported by the DOE NNSA Steward Science Graduate Fellowship under grant number DE-NA0003864. The methods for this work are implemented using the Python programming language\footnote{www.python.org}, and make extensive use of the NumPy/SciPy library \citep{NumPyArray, SciPy}.




\bibliographystyle{mnras}
\bibliography{references} 

\begin{thebibliography}{}
\makeatletter
\relax
\def\mn@urlcharsother{\let\do\@makeother \do\$\do\&\do\#\do\^\do\_\do\%\do\~}
\def\mn@doi{\begingroup\mn@urlcharsother \@ifnextchar [ {\mn@doi@}
  {\mn@doi@[]}}
\def\mn@doi@[#1]#2{\def\@tempa{#1}\ifx\@tempa\@empty \href
  {http://dx.doi.org/#2} {doi:#2}\else \href {http://dx.doi.org/#2} {#1}\fi
  \endgroup}
\def\mn@eprint#1#2{\mn@eprint@#1:#2::\@nil}
\def\mn@eprint@arXiv#1{\href {http://arxiv.org/abs/#1} {{\tt arXiv:#1}}}
\def\mn@eprint@dblp#1{\href {http://dblp.uni-trier.de/rec/bibtex/#1.xml}
  {dblp:#1}}
\def\mn@eprint@#1:#2:#3:#4\@nil{\def\@tempa {#1}\def\@tempb {#2}\def\@tempc
  {#3}\ifx \@tempc \@empty \let \@tempc \@tempb \let \@tempb \@tempa \fi \ifx
  \@tempb \@empty \def\@tempb {arXiv}\fi \@ifundefined
  {mn@eprint@\@tempb}{\@tempb:\@tempc}{\expandafter \expandafter \csname
  mn@eprint@\@tempb\endcsname \expandafter{\@tempc}}}

\bibitem[\protect\citeauthoryear{{Abbott} et~al.,}{{Abbott}
  et~al.}{2017a}]{2017PhRvL.119p1101A}
{Abbott} B.~P.,  et~al., 2017a, \mn@doi [Physical Review Letters]
  {10.1103/PhysRevLett.119.161101}, \href
  {http://adsabs.harvard.edu/abs/2017PhRvL.119p1101A} {119, 161101}

\bibitem[\protect\citeauthoryear{{Abbott} et~al.,}{{Abbott}
  et~al.}{2017b}]{2017Natur.551...85A}
{Abbott} B.~P.,  et~al., 2017b, \mn@doi [\nat] {10.1038/nature24471}, \href
  {http://adsabs.harvard.edu/abs/2017Natur.551...85A} {551, 85}

\bibitem[\protect\citeauthoryear{{Abbott} et~al.,}{{Abbott}
  et~al.}{2017c}]{2017ApJ...848L..12A}
{Abbott} B.~P.,  et~al., 2017c, \mn@doi [\apjl] {10.3847/2041-8213/aa91c9},
  \href {http://adsabs.harvard.edu/abs/2017ApJ...848L..12A} {848, L12}

\bibitem[\protect\citeauthoryear{{Abbott} et~al.,}{{Abbott}
  et~al.}{2017d}]{2017ApJ...848L..13A}
{Abbott} B.~P.,  et~al., 2017d, \mn@doi [\apjl] {10.3847/2041-8213/aa920c},
  \href {http://adsabs.harvard.edu/abs/2017ApJ...848L..13A} {848, L13}

\bibitem[\protect\citeauthoryear{{Alam} et~al.,}{{Alam}
  et~al.}{2015}]{2015ApJS..219...12A}
{Alam} S.,  et~al., 2015, \mn@doi [\apjs] {10.1088/0067-0049/219/1/12}, \href
  {http://adsabs.harvard.edu/abs/2015ApJS..219...12A} {219, 12}

\bibitem[\protect\citeauthoryear{{Amaro-Seoane} et~al.,}{{Amaro-Seoane}
  et~al.}{2017}]{2017arXiv170200786A}
{Amaro-Seoane} P.,  et~al., 2017, preprint, \href
  {http://adsabs.harvard.edu/abs/2017arXiv170200786A} {} (\mn@eprint {arXiv}
  {1702.00786})

\bibitem[\protect\citeauthoryear{{Anholm}, {Ballmer}, {Creighton}, {Price}  \&
  {Siemens}}{{Anholm} et~al.}{2009}]{anholm09}
{Anholm} M.,  {Ballmer} S.,  {Creighton} J.~D.~E.,  {Price} L.~R.,   {Siemens}
  X.,  2009, \mn@doi [Phys. Rev. D] {10.1103/PhysRevD.79.084030}, \href
  {http://adsabs.harvard.edu/abs/2009PhRvD..79h4030A} {79, 084030}

\bibitem[\protect\citeauthoryear{{Arzoumanian} et~al.,}{{Arzoumanian}
  et~al.}{2018}]{2018ApJ...859...47A}
{Arzoumanian} Z.,  et~al., 2018, \mn@doi [\apj] {10.3847/1538-4357/aabd3b},
  \href {http://adsabs.harvard.edu/abs/2018ApJ...859...47A} {859, 47}

\bibitem[\protect\citeauthoryear{{Babak} \& {Sesana}}{{Babak} \&
  {Sesana}}{2012}]{2012PhRvD..85d4034B}
{Babak} S.,  {Sesana} A.,  2012, \mn@doi [\prd] {10.1103/PhysRevD.85.044034},
  \href {http://adsabs.harvard.edu/abs/2012PhRvD..85d4034B} {85, 044034}

\bibitem[\protect\citeauthoryear{Babak et~al.,}{Babak
  et~al.}{2016}]{doi:10.1093/mnras/stv2092}
Babak S.,  et~al., 2016, \mn@doi [Monthly Notices of the Royal Astronomical
  Society] {10.1093/mnras/stv2092}, 455, 1665

\bibitem[\protect\citeauthoryear{{Boyle} \& {Pen}}{{Boyle} \&
  {Pen}}{2012}]{2012PhRvD..86l4028B}
{Boyle} L.,  {Pen} U.-L.,  2012, \mn@doi [\prd] {10.1103/PhysRevD.86.124028},
  \href {http://adsabs.harvard.edu/abs/2012PhRvD..86l4028B} {86, 124028}

\bibitem[\protect\citeauthoryear{{Burke-Spolaor}}{{Burke-Spolaor}}{2013}]{2013CQGra..30v4013B}
{Burke-Spolaor} S.,  2013, \mn@doi [Classical and Quantum Gravity]
  {10.1088/0264-9381/30/22/224013}, \href
  {http://adsabs.harvard.edu/abs/2013CQGra..30v4013B} {30, 224013}

\bibitem[\protect\citeauthoryear{{Burt}, {Lommen}  \& {Finn}}{{Burt}
  et~al.}{2011}]{2011ApJ...730...17B}
{Burt} B.~J.,  {Lommen} A.~N.,   {Finn} L.~S.,  2011, \mn@doi [\apj]
  {10.1088/0004-637X/730/1/17}, \href
  {https://ui.adsabs.harvard.edu/\#abs/2011ApJ...730...17B} {730, 17}

\bibitem[\protect\citeauthoryear{{Chornock} et~al.,}{{Chornock}
  et~al.}{2017}]{2017ApJ...848L..19C}
{Chornock} R.,  et~al., 2017, \mn@doi [\apjl] {10.3847/2041-8213/aa905c}, \href
  {http://adsabs.harvard.edu/abs/2017ApJ...848L..19C} {848, L19}

\bibitem[\protect\citeauthoryear{{Dotti}, {Sesana}  \& {Decarli}}{{Dotti}
  et~al.}{2012}]{2012AdAst2012E...3D}
{Dotti} M.,  {Sesana} A.,   {Decarli} R.,  2012, \mn@doi [Advances in
  Astronomy] {10.1155/2012/940568}, \href
  {http://adsabs.harvard.edu/abs/2012AdAst2012E...3D} {2012, 940568}

\bibitem[\protect\citeauthoryear{{Duncan} et~al.,}{{Duncan}
  et~al.}{2014}]{2014MNRAS.444.2960D}
{Duncan} K.,  et~al., 2014, \mn@doi [\mnras] {10.1093/mnras/stu1622}, \href
  {http://adsabs.harvard.edu/abs/2014MNRAS.444.2960D} {444, 2960}

\bibitem[\protect\citeauthoryear{Fishbach, Gray, Hernandez, Qi, Sur  \& {and
  members of the LIGO Scientific Collaboration and the Virgo
  Collaboration}}{Fishbach et~al.}{2018}]{Fishbach:2018gjp}
Fishbach M.,  Gray R.,  Hernandez I.~M.,  Qi H.,  Sur A.,   {and members of the
  LIGO Scientific Collaboration and the Virgo Collaboration} 2018, preprint
  (\mn@eprint {arXiv} {1807.05667})

\bibitem[\protect\citeauthoryear{{Gaia Collaboration} et~al.,}{{Gaia
  Collaboration} et~al.}{2016}]{2016A&A...595A...1G}
{Gaia Collaboration} et~al., 2016, \mn@doi [\aap]
  {10.1051/0004-6361/201629272}, \href
  {http://adsabs.harvard.edu/abs/2016A%26A...595A...1G} {595, A1}

\bibitem[\protect\citeauthoryear{{Gardner} et~al.,}{{Gardner}
  et~al.}{2006}]{2006SSRv..123..485G}
{Gardner} J.~P.,  et~al., 2006, \mn@doi [\ssr] {10.1007/s11214-006-8315-7},
  \href {http://adsabs.harvard.edu/abs/2006SSRv..123..485G} {123, 485}

\bibitem[\protect\citeauthoryear{{Gilmozzi} \& {Spyromilio}}{{Gilmozzi} \&
  {Spyromilio}}{2007}]{2007Msngr.127...11G}
{Gilmozzi} R.,  {Spyromilio} J.,  2007, The Messenger, \href
  {http://adsabs.harvard.edu/abs/2007Msngr.127...11G} {127}

\bibitem[\protect\citeauthoryear{Goldstein, Veitch, Sesana  \&
  Vecchio}{Goldstein et~al.}{2018}]{Goldstein:2017qub}
Goldstein J.,  Veitch J.,  Sesana A.,   Vecchio A.,  2018, \mn@doi [Mon. Not.
  Roy. Astron. Soc.] {10.1093/mnras/sty892}, 477, 5447

\bibitem[\protect\citeauthoryear{{G{\'o}rski}, {Hivon}, {Banday}, {Wandelt},
  {Hansen}, {Reinecke}  \& {Bartelmann}}{{G{\'o}rski} et~al.}{2005}]{HEALPix}
{G{\'o}rski} K.~M.,  {Hivon} E.,  {Banday} A.~J.,  {Wandelt} B.~D.,  {Hansen}
  F.~K.,  {Reinecke} M.,   {Bartelmann} M.,  2005, \mn@doi [\apj]
  {10.1086/427976}, \href {http://adsabs.harvard.edu/abs/2005ApJ...622..759G}
  {622, 759}

\bibitem[\protect\citeauthoryear{{Guo} et~al.,}{{Guo}
  et~al.}{2011}]{2011MNRAS.413..101G}
{Guo} Q.,  et~al., 2011, \mn@doi [\mnras] {10.1111/j.1365-2966.2010.18114.x},
  \href {http://adsabs.harvard.edu/abs/2011MNRAS.413..101G} {413, 101}

\bibitem[\protect\citeauthoryear{Haiman}{Haiman}{2017}]{haiman_electromagnetic_2017}
Haiman Z.,  2017, \mn@doi [Physical Review D] {10.1103/PhysRevD.96.023004}, 96

\bibitem[\protect\citeauthoryear{{Hazboun} \& {Larson}}{{Hazboun} \&
  {Larson}}{2016}]{2016arXiv160703459H}
{Hazboun} J.~S.,  {Larson} S.~L.,  2016, preprint, \href
  {http://adsabs.harvard.edu/abs/2016arXiv160703459H} {} (\mn@eprint {arXiv}
  {1607.03459})

\bibitem[\protect\citeauthoryear{{Henriques}, {White}, {Lemson}, {Thomas},
  {Guo}, {Marleau}  \& {Overzier}}{{Henriques}
  et~al.}{2012}]{2012MNRAS.421.2904H}
{Henriques} B.~M.~B.,  {White} S.~D.~M.,  {Lemson} G.,  {Thomas} P.~A.,  {Guo}
  Q.,  {Marleau} G.-D.,   {Overzier} R.~A.,  2012, \mn@doi [\mnras]
  {10.1111/j.1365-2966.2012.20521.x}, \href
  {http://adsabs.harvard.edu/abs/2012MNRAS.421.2904H} {421, 2904}

\bibitem[\protect\citeauthoryear{Holgado, Sesana, Sandrinelli, Covino, Treves,
  Liu  \& Ricker}{Holgado et~al.}{2018}]{holgado_pulsar_2018}
Holgado A.~M.,  Sesana A.,  Sandrinelli A.,  Covino S.,  Treves A.,  Liu X.,
  Ricker P.,  2018, \mn@doi [MNRAS Letters] {10.1093/mnrasl/sly158}, 481, L74

\bibitem[\protect\citeauthoryear{{Inayoshi}, {Ichikawa}  \&
  {Haiman}}{{Inayoshi} et~al.}{2018}]{2018ApJ...863L..36I}
{Inayoshi} K.,  {Ichikawa} K.,   {Haiman} Z.,  2018, \mn@doi [\apjl]
  {10.3847/2041-8213/aad8ad}, \href
  {http://adsabs.harvard.edu/abs/2018ApJ...863L..36I} {863, L36}

\bibitem[\protect\citeauthoryear{{Jaranowski}, {Kr{\'o}lak}  \&
  {Schutz}}{{Jaranowski} et~al.}{1998}]{1998PhRvD..58f3001J}
{Jaranowski} P.,  {Kr{\'o}lak} A.,   {Schutz} B.~F.,  1998, \mn@doi [\prd]
  {10.1103/PhysRevD.58.063001}, \href
  {http://adsabs.harvard.edu/abs/1998PhRvD..58f3001J} {58, 063001}

\bibitem[\protect\citeauthoryear{Jones, Oliphant, Peterson  et~al.}{Jones
  et~al.}{01  }]{SciPy}
Jones E.,  Oliphant T.,  Peterson P.,   et~al., 2001--, {SciPy}: Open source
  scientific tools for {Python}, \url {http://www.scipy.org/}

\bibitem[\protect\citeauthoryear{{Kaiser} et~al.,}{{Kaiser}
  et~al.}{2002}]{2002SPIE.4836..154K}
{Kaiser} N.,  et~al., 2002, in {Tyson} J.~A.,  {Wolff} S.,  eds,  \procspie
  Vol. 4836, Survey and Other Telescope Technologies and Discoveries. pp
  154--164, \mn@doi{10.1117/12.457365}

\bibitem[\protect\citeauthoryear{{Kelley}, {Blecha}  \& {Hernquist}}{{Kelley}
  et~al.}{2017}]{2017MNRAS.464.3131K}
{Kelley} L.~Z.,  {Blecha} L.,   {Hernquist} L.,  2017, \mn@doi [\mnras]
  {10.1093/mnras/stw2452}, \href
  {http://adsabs.harvard.edu/abs/2017MNRAS.464.3131K} {464, 3131}

\bibitem[\protect\citeauthoryear{{Kelley}, {Blecha}, {Hernquist}, {Sesana}  \&
  {Taylor}}{{Kelley} et~al.}{2018}]{2018MNRAS.477..964K}
{Kelley} L.~Z.,  {Blecha} L.,  {Hernquist} L.,  {Sesana} A.,   {Taylor} S.~R.,
  2018, \mn@doi [\mnras] {10.1093/mnras/sty689}, \href
  {https://ui.adsabs.harvard.edu/\#abs/2018MNRAS.477..964K} {477, 964}

\bibitem[\protect\citeauthoryear{{Klein} et~al.,}{{Klein}
  et~al.}{2016}]{2016PhRvD..93b4003K}
{Klein} A.,  et~al., 2016, \mn@doi [\prd] {10.1103/PhysRevD.93.024003}, \href
  {http://adsabs.harvard.edu/abs/2016PhRvD..93b4003K} {93, 024003}

\bibitem[\protect\citeauthoryear{Kormendy \& Ho}{Kormendy \& Ho}{2013}]{korho}
Kormendy J.,  Ho L.~C.,  2013, \mn@doi [Annual Review of Astronomy and
  Astrophysics] {10.1146/annurev-astro-082708-101811}, 51, 511

\bibitem[\protect\citeauthoryear{{Koss} et~al.,}{{Koss}
  et~al.}{2018}]{2018Natur.563..214K}
{Koss} M.~J.,  et~al., 2018, \mn@doi [\nat] {10.1038/s41586-018-0652-7}, \href
  {http://adsabs.harvard.edu/abs/2018Natur.563..214K} {563, 214}

\bibitem[\protect\citeauthoryear{{LSST Science Collaboration} et~al.,}{{LSST
  Science Collaboration} et~al.}{2009}]{2009arXiv0912.0201L}
{LSST Science Collaboration} et~al., 2009, arXiv e-prints, \href
  {http://adsabs.harvard.edu/abs/2009arXiv0912.0201L} {}

\bibitem[\protect\citeauthoryear{{Lam}}{{Lam}}{2018}]{2018ApJ...868...33L}
{Lam} M.~T.,  2018, \mn@doi [\apj] {10.3847/1538-4357/aae533}, \href
  {https://ui.adsabs.harvard.edu/\#abs/2018ApJ...868...33L} {868, 33}

\bibitem[\protect\citeauthoryear{{Lentati} et~al.,}{{Lentati}
  et~al.}{2015}]{2015MNRAS.453.2576L}
{Lentati} L.,  et~al., 2015, \mn@doi [\mnras] {10.1093/mnras/stv1538}, \href
  {http://adsabs.harvard.edu/abs/2015MNRAS.453.2576L} {453, 2576}

\bibitem[\protect\citeauthoryear{{Longhetti} \& {Saracco}}{{Longhetti} \&
  {Saracco}}{2009}]{2009MNRAS.394..774L}
{Longhetti} M.,  {Saracco} P.,  2009, \mn@doi [\mnras]
  {10.1111/j.1365-2966.2008.14375.x}, \href
  {http://adsabs.harvard.edu/abs/2009MNRAS.394..774L} {394, 774}

\bibitem[\protect\citeauthoryear{{Lotz}, {Jonsson}, {Cox}  \& {Primack}}{{Lotz}
  et~al.}{2008}]{2008MNRAS.391.1137L}
{Lotz} J.~M.,  {Jonsson} P.,  {Cox} T.~J.,   {Primack} J.~R.,  2008, \mn@doi
  [\mnras] {10.1111/j.1365-2966.2008.14004.x}, \href
  {http://adsabs.harvard.edu/abs/2008MNRAS.391.1137L} {391, 1137}

\bibitem[\protect\citeauthoryear{{McAlpine} et~al.,}{{McAlpine}
  et~al.}{2016}]{2016A&C....15...72M}
{McAlpine} S.,  et~al., 2016, \mn@doi [Astronomy and Computing]
  {10.1016/j.ascom.2016.02.004}, \href
  {http://adsabs.harvard.edu/abs/2016A%26C....15...72M} {15, 72}

\bibitem[\protect\citeauthoryear{{Meiron} \& {Laor}}{{Meiron} \&
  {Laor}}{2013}]{2013MNRAS.433.2502M}
{Meiron} Y.,  {Laor} A.,  2013, \mn@doi [\mnras] {10.1093/mnras/stt922}, \href
  {http://adsabs.harvard.edu/abs/2013MNRAS.433.2502M} {433, 2502}

\bibitem[\protect\citeauthoryear{{Merloni} et~al.,}{{Merloni}
  et~al.}{2012}]{2012arXiv1209.3114M}
{Merloni} A.,  et~al., 2012, arXiv e-prints, \href
  {http://adsabs.harvard.edu/abs/2012arXiv1209.3114M} {}

\bibitem[\protect\citeauthoryear{{Middleton}, {Chen}, {Del Pozzo}, {Sesana}  \&
  {Vecchio}}{{Middleton} et~al.}{2018}]{2018NatCo...9..573M}
{Middleton} H.,  {Chen} S.,  {Del Pozzo} W.,  {Sesana} A.,   {Vecchio} A.,
  2018, \mn@doi [Nature Communications] {10.1038/s41467-018-02916-7}, \href
  {http://adsabs.harvard.edu/abs/2018NatCo...9..573M} {9, 573}

\bibitem[\protect\citeauthoryear{{Mingarelli} et~al.,}{{Mingarelli}
  et~al.}{2017}]{2017NatAs...1..886M}
{Mingarelli} C. M.~F.,  et~al., 2017, \mn@doi [Nature Astronomy]
  {10.1038/s41550-017-0299-6}, \href
  {https://ui.adsabs.harvard.edu/\#abs/2017NatAs...1..886M} {1, 886}

\bibitem[\protect\citeauthoryear{{Nandra} et~al.,}{{Nandra}
  et~al.}{2013}]{2013arXiv1306.2307N}
{Nandra} K.,  et~al., 2013, arXiv e-prints, \href
  {http://adsabs.harvard.edu/abs/2013arXiv1306.2307N} {}

\bibitem[\protect\citeauthoryear{{Phinney}}{{Phinney}}{2001}]{Phinney2001}
{Phinney} E.~S.,  2001, arXiv Astrophysics e-prints, \href
  {http://adsabs.harvard.edu/abs/2001astro.ph..8028P} {}

\bibitem[\protect\citeauthoryear{{Planck Collaboration} et~al.,}{{Planck
  Collaboration} et~al.}{2016}]{2016A&A...594A..13P}
{Planck Collaboration} et~al., 2016, \mn@doi [\aap]
  {10.1051/0004-6361/201525830}, \href
  {http://adsabs.harvard.edu/abs/2016A%26A...594A..13P} {594, A13}

\bibitem[\protect\citeauthoryear{{Ravi}, {Wyithe}, {Hobbs}, {Shannon},
  {Manchester}, {Yardley}  \& {Keith}}{{Ravi}
  et~al.}{2012}]{2012ApJ...761...84R}
{Ravi} V.,  {Wyithe} J.~S.~B.,  {Hobbs} G.,  {Shannon} R.~M.,  {Manchester}
  R.~N.,  {Yardley} D.~R.~B.,   {Keith} M.~J.,  2012, \mn@doi [\apj]
  {10.1088/0004-637X/761/2/84}, \href
  {http://adsabs.harvard.edu/abs/2012ApJ...761...84R} {761, 84}

\bibitem[\protect\citeauthoryear{{Ravi}, {Wyithe}, {Shannon}  \&
  {Hobbs}}{{Ravi} et~al.}{2015}]{2015MNRAS.447.2772R}
{Ravi} V.,  {Wyithe} J.~S.~B.,  {Shannon} R.~M.,   {Hobbs} G.,  2015, \mn@doi
  [\mnras] {10.1093/mnras/stu2659}, \href
  {http://adsabs.harvard.edu/abs/2015MNRAS.447.2772R} {447, 2772}

\bibitem[\protect\citeauthoryear{{Rosado} \& {Sesana}}{{Rosado} \&
  {Sesana}}{2014}]{2014MNRAS.439.3986R}
{Rosado} P.~A.,  {Sesana} A.,  2014, \mn@doi [\mnras] {10.1093/mnras/stu254},
  \href {http://adsabs.harvard.edu/abs/2014MNRAS.439.3986R} {439, 3986}

\bibitem[\protect\citeauthoryear{{Rosado}, {Sesana}  \& {Gair}}{{Rosado}
  et~al.}{2015}]{2015MNRAS.451.2417R}
{Rosado} P.~A.,  {Sesana} A.,   {Gair} J.,  2015, \mn@doi [\mnras]
  {10.1093/mnras/stv1098}, \href
  {http://adsabs.harvard.edu/abs/2015MNRAS.451.2417R} {451, 2417}

\bibitem[\protect\citeauthoryear{Schutz}{Schutz}{1986}]{schutz_determining_1986}
Schutz B.~F.,  1986, \mn@doi [Nature] {10.1038/323310a0}, 323, 310

\bibitem[\protect\citeauthoryear{{Sesana}}{{Sesana}}{2013}]{2013MNRAS.433L...1S}
{Sesana} A.,  2013, \mn@doi [\mnras] {10.1093/mnrasl/slt034}, \href
  {http://adsabs.harvard.edu/abs/2013MNRAS.433L...1S} {433, L1}

\bibitem[\protect\citeauthoryear{{Sesana} \& {Vecchio}}{{Sesana} \&
  {Vecchio}}{2010}]{2010PhRvD..81j4008S}
{Sesana} A.,  {Vecchio} A.,  2010, \mn@doi [\prd] {10.1103/PhysRevD.81.104008},
  \href {http://adsabs.harvard.edu/abs/2010PhRvD..81j4008S} {81, 104008}

\bibitem[\protect\citeauthoryear{{Sesana}, {Vecchio}  \& {Colacino}}{{Sesana}
  et~al.}{2008}]{2008MNRAS.390..192S}
{Sesana} A.,  {Vecchio} A.,   {Colacino} C.~N.,  2008, \mn@doi [\mnras]
  {10.1111/j.1365-2966.2008.13682.x}, \href
  {http://adsabs.harvard.edu/abs/2008MNRAS.390..192S} {390, 192}

\bibitem[\protect\citeauthoryear{{Sesana}, {Vecchio}  \& {Volonteri}}{{Sesana}
  et~al.}{2009}]{2009MNRAS.394.2255S}
{Sesana} A.,  {Vecchio} A.,   {Volonteri} M.,  2009, \mn@doi [\mnras]
  {10.1111/j.1365-2966.2009.14499.x}, \href
  {http://adsabs.harvard.edu/abs/2009MNRAS.394.2255S} {394, 2255}

\bibitem[\protect\citeauthoryear{{Sesana}, {Roedig}, {Reynolds}  \&
  {Dotti}}{{Sesana} et~al.}{2012}]{2012MNRAS.420..860S}
{Sesana} A.,  {Roedig} C.,  {Reynolds} M.~T.,   {Dotti} M.,  2012, \mn@doi
  [\mnras] {10.1111/j.1365-2966.2011.20097.x}, \href
  {http://adsabs.harvard.edu/abs/2012MNRAS.420..860S} {420, 860}

\bibitem[\protect\citeauthoryear{{Sesana}, {Shankar}, {Bernardi}  \&
  {Sheth}}{{Sesana} et~al.}{2016}]{2016MNRAS.463L...6S}
{Sesana} A.,  {Shankar} F.,  {Bernardi} M.,   {Sheth} R.~K.,  2016, \mn@doi
  [\mnras] {10.1093/mnrasl/slw139}, \href
  {http://adsabs.harvard.edu/abs/2016MNRAS.463L...6S} {463, L6}

\bibitem[\protect\citeauthoryear{{Sesana}, {Haiman}, {Kocsis}  \&
  {Kelley}}{{Sesana} et~al.}{2018}]{2018ApJ...856...42S}
{Sesana} A.,  {Haiman} Z.,  {Kocsis} B.,   {Kelley} L.~Z.,  2018, \mn@doi
  [\apj] {10.3847/1538-4357/aaad0f}, \href
  {http://adsabs.harvard.edu/abs/2018ApJ...856...42S} {856, 42}

\bibitem[\protect\citeauthoryear{{Shankar} et~al.,}{{Shankar}
  et~al.}{2016}]{shankar}
{Shankar} F.,  et~al., 2016, \mn@doi [\mnras] {10.1093/mnras/stw678}, \href
  {http://adsabs.harvard.edu/abs/2016MNRAS.460.3119S} {460, 3119}

\bibitem[\protect\citeauthoryear{{Shannon} et~al.,}{{Shannon}
  et~al.}{2015}]{2015Sci...349.1522S}
{Shannon} R.~M.,  et~al., 2015, \mn@doi [Science] {10.1126/science.aab1910},
  \href {http://adsabs.harvard.edu/abs/2015Sci...349.1522S} {349, 1522}

\bibitem[\protect\citeauthoryear{{Shapiro}, {Bacon}, {Hendry}  \&
  {Hoyle}}{{Shapiro} et~al.}{2010}]{2010MNRAS.404..858S}
{Shapiro} C.,  {Bacon} D.~J.,  {Hendry} M.,   {Hoyle} B.,  2010, \mn@doi
  [\mnras] {10.1111/j.1365-2966.2010.16317.x}, \href
  {http://adsabs.harvard.edu/abs/2010MNRAS.404..858S} {404, 858}

\bibitem[\protect\citeauthoryear{{Simon}, {Polin}, {Lommen}, {Stappers},
  {Finn}, {Jenet}  \& {Christy}}{{Simon} et~al.}{2014}]{2014ApJ...784...60S}
{Simon} J.,  {Polin} A.,  {Lommen} A.,  {Stappers} B.,  {Finn} L.~S.,  {Jenet}
  F.~A.,   {Christy} B.,  2014, \mn@doi [\apj] {10.1088/0004-637X/784/1/60},
  \href {https://ui.adsabs.harvard.edu/\#abs/2014ApJ...784...60S} {784, 60}

\bibitem[\protect\citeauthoryear{{Springel} et~al.,}{{Springel}
  et~al.}{2005}]{2005Natur.435..629S}
{Springel} V.,  et~al., 2005, \mn@doi [\nat] {10.1038/nature03597}, \href
  {http://adsabs.harvard.edu/abs/2005Natur.435..629S} {435, 629}

\bibitem[\protect\citeauthoryear{Tanaka, Menou  \& Haiman}{Tanaka
  et~al.}{2012}]{tanaka_electromagnetic_2012}
Tanaka T.,  Menou K.,   Haiman Z.,  2012, \mn@doi [MNRAS]
  {10.1111/j.1365-2966.2011.20083.x}, 420, 705

\bibitem[\protect\citeauthoryear{Tang, MacFadyen  \& Haiman}{Tang
  et~al.}{2017}]{doi:10.1093/mnras/stx1130}
Tang Y.,  MacFadyen A.,   Haiman Z.,  2017, \mn@doi [Monthly Notices of the
  Royal Astronomical Society] {10.1093/mnras/stx1130}, 469, 4258

\bibitem[\protect\citeauthoryear{{Verbiest} et~al.,}{{Verbiest}
  et~al.}{2016}]{2016MNRAS.458.1267V}
{Verbiest} J.~P.~W.,  et~al., 2016, \mn@doi [\mnras] {10.1093/mnras/stw347},
  \href {http://adsabs.harvard.edu/abs/2016MNRAS.458.1267V} {458, 1267}

\bibitem[\protect\citeauthoryear{{Vogelsberger} et~al.,}{{Vogelsberger}
  et~al.}{2014}]{2014Natur.509..177V}
{Vogelsberger} M.,  et~al., 2014, \mn@doi [\nat] {10.1038/nature13316}, \href
  {http://adsabs.harvard.edu/abs/2014Natur.509..177V} {509, 177}

\bibitem[\protect\citeauthoryear{{Zhu} et~al.,}{{Zhu}
  et~al.}{2015}]{2015MNRAS.449.1650Z}
{Zhu} X.-J.,  et~al., 2015, \mn@doi [\mnras] {10.1093/mnras/stv381}, \href
  {http://adsabs.harvard.edu/abs/2015MNRAS.449.1650Z} {449, 1650}

\bibitem[\protect\citeauthoryear{van~der Walt, Colbert  \& Varoquaux}{van~der
  Walt et~al.}{2011}]{NumPyArray}
van~der Walt S.,  Colbert S.~C.,   Varoquaux G.,  2011, CoRR, abs/1102.1523

\makeatother
\end{thebibliography}



\appendix


\bsp	
\label{lastpage}
\end{document}